\documentclass[lettersize,journal]{IEEEtran}
\usepackage{amsmath}
\usepackage{amsthm}
\usepackage{amsfonts}
\usepackage{amssymb}
\usepackage{url}
\usepackage{framed}
\usepackage{caption}
\usepackage{color}
\usepackage{chemfig}
\usepackage[version=4]{mhchem}
\usepackage{graphicx}
\usepackage{float}

\begin{document}

\title{Hijacking Living Cells with Surface Engineering for the Internet of Bio-Nano Things}
\author{Ekin Ince
        and Murat Kuscu
       \thanks{The authors are with the Nano/Bio/Physical Information and Communications Laboratory (CALICO Lab), Department of Electrical and Electronics Engineering, Koç University, Istanbul, Turkey (e-mail: \{ekince23, mkuscu\}@ku.edu.tr).}
	   \thanks{This work was supported in part by the European
Union’s Horizon 2020 Research and Innovation Programme through the Marie Skłodowska-Curie Individual Fellowship under Grant Agreement 101028935, and by The Scientific and Technological Research Council of Turkey (TUBITAK) under Grants \#123E516 and \#123C592.}}



\maketitle

\begin{abstract}
The Internet of Bio-Nano Things (IoBNT) promises to revolutionize healthcare by interfacing the cyber domain with the living systems at unprecedented resolution. Realizing this vision hinges on the development of Bio-Nano Things (BNTs), i.e., functional nodes capable of sensing, actuation, and communications within biological environments. Existing BNT architectures, e.g., nanomaterial-based, biosynthetic, and passive molecular agents, face significant limitations, including toxicity, lack of autonomy, or the safety and metabolic burdens associated with genetic modification. This paper posits a fourth paradigm: the transient hijacking of living cells via non-genetic cell surface engineering (NG-CSE) to enable living BNTs. NGCSE allows for the precise, reversible functionalization of cell membranes with synthetic molecular machinery, reprogramming cellular functions and interactions without altering the genome. It uniquely combines the inherent biocompatibility and agency of living cells with the programmability enabled by nanotechnology, mitigating the risks of genetic engineering. We critically review the toolbox of NG-CSE and explore the opportunities it unlocks for IoBNT, including programmable cell-cell communication, dynamic network topologies, and improved bio-cyber interfacing. Moreover, we propose novel IoBNT architectures that leverage these capabilities, such as circulating sentinel networks exploiting cellular agency for continuous liquid biopsy, and rationally designed, in vitro biocomputers exploiting interkingdom interactions. We also outline the critical challenges in modeling and exploiting cellular agency with NG-CSE, providing a roadmap for the effective utilization of NG-CSE-enabled living BNTs within IoBNT.

\end{abstract}

\begin{IEEEkeywords}
Internet of Bio-Nano Things, non-genetic cell surface engineering, bio-cyber systems, molecular communications
\end{IEEEkeywords}
\section{Introduction}\label{Int}

\IEEEPARstart{T}{he} Internet of Bio-Nano Things (IoBNT) is an emerging networking framework poised to dissolve the boundaries between the intricate biochemical domain of living systems and the digital cyber world \cite{akyildiz2015internet}. Unlike the conventional Internet of Things, which connects macroscopic electronic devices, IoBNT is envisioned as a heterogeneous network of living, biological entities and engineered synthetic devices, collectively termed Bio-Nano Things (BNTs) \cite{kuscu2021internet}. These BNTs operate within unconventional environments, such as the human body, and communicate mainly through molecular communications (MC) rather than electromagnetic waves \cite{akan2017fundamentals}. The objective of this framework is to establish seamless, real-time bio-cyber interfaces capable of sensing and controlling biological dynamics with unprecedented molecular precision and high spatiotemporal resolution. The realization of IoBNT, therefore, promises paradigm-shifting applications, most notably in personalized medicine, enabling continuous intra-body health monitoring, closed-loop theranostics, and ultra-precise therapeutic interventions \cite{kuscu2021internet, sun2025sensing}.

The functional capacity of the IoBNT framework, however, is fundamentally constrained by the capabilities of the BNTs themselves, i.e., the nodes responsible for sensing, actuation, computation, and communication at the molecular level. Research in IoBNT to date has largely advanced along three distinct BNT paradigms, each with inherent strengths and critical limitations. (i) Nanomaterial-based abiotic devices, such as graphene sensors or magnetic nanorobots, provide sophisticated functionality analogous to macroscopic electronics but suffer from poor biocompatibility, biofouling, potential toxicity, and a dependency on external power sources \cite{civas2023graphene, kuscu2021fabrication}. (ii) Passive molecular agents, such as quantum dots, fluorescent molecules, or enzyme-powered nanoparticles, exhibit better biocompatibility but lack autonomy, programmability, and active control, relegating them to simple, emergent communication tasks that are unsuitable for advanced applications \cite{kuscu2015internet, von2011nanoparticles}. (iii) Biosynthetic devices, which use genetically engineered cells as BNTs, harness the complexity of living systems to perform advanced functions \cite{koca2024bacterial, unluturk2015genetically}. However, this approach is fraught with significant biosafety concerns, including the potential for off-target genetic effects, horizontal gene transfer, and adverse immune responses. Furthermore, synthetic genetic circuits impose a substantial metabolic burden on the host cell, which can compromise its viability and lead to unpredictable behavior \emph{in vivo} \cite{wu2016metabolic}. 

The current IoBNT landscape, therefore, has a critical gap, which is the need for a BNT architecture that combines the programmability of synthetic devices with the biocompatibility and autonomous agency of living systems, while circumventing the liabilities associated with both foreign nanomaterials and genetic modification. To address this unmet need, this paper proposes a fourth BNT paradigm for IoBNT, which is the utilization of living cells transiently hijacked through \emph{non-genetic cell surface engineering (NG-CSE)}.

NG-CSE is an emerging biotechnology that involves the precise, external modification of the cell's plasma membrane with synthetic functional units, such as aptamers, DNA nanostructures, nanoparticles, or molecular logic circuits, without altering the cell's underlying genome \cite{grupi2019interfacing, abbina2017surface}. It allows living cells to be rapidly and reversibly repurposed as controllable, responsive, functional devices. Importantly, NG-CSE preserves the cell's native metabolism and viability while endowing it with programmable interfaces for engineered sensing, communication, and actuation. It uniquely merges the agency of living systems, i.e., their ability to autonomously navigate, adapt, and self-regulate, with the rational design principles of nanotechnology.

In this paper, we posit that NG-CSE can be a foundational strategy for the future of IoBNT. We argue that by transforming the cell membrane into a programmable bio-cyber interface, NG-CSE overcomes the limitations of existing BNT paradigms and unlocks novel network architectures that explicitly exploit, rather than suppress, cellular agency. We provide a comprehensive analysis of the NG-CSE toolbox, explore the unique opportunities it enables for IoBNT, and envision novel IoBNT applications that leverage these capabilities. Moreover, we delineate the key engineering and translational challenges, providing a research roadmap for the effective merge of these two emerging fields.

The remainder of this paper is organized as follows. Section II provides an overview of the IoBNT framework and critically reviews the existing BNT architectures, solidifying the motivation for NG-CSE-enabled living BNTs. Section III details the toolbox of materials and methods used for NG-CSE. Section IV explores the key opportunities and capabilities that NG-CSE enables for IoBNT. Section V presents potential IoBNT applications uniquely enabled by this new paradigm. Section VI discusses the critical challenges and future research directions, focusing on modeling cellular agency and managing BNT lifecycles. Finally, Section VII concludes the paper.

\section{Internet of Bio-Nano Things}
\subsection{Background and Motivation}
IoBNT is an unconventional networking framework that extends the foundational principles of the Internet of Things (IoT) into the biochemical domain of living systems \cite{akyildiz2015internet}. It is envisioned as a heterogeneous network of micro- and nanoscale functional devices and biological entities. These BNTs are designed to operate in complex biological environments, most notably the human body, and mainly communicate via the exchange of molecules. 

The main objective of IoBNT is to establish a seamless, real-time interface between the biological and cyber domains, enabling the sensing and control of biological dynamics with unprecedented spatiotemporal resolution. This bio-cyber integration promises paradigm-shifting applications across various fields. In healthcare, IoBNT aims to revolutionize personalized medicine through continuous intrabody health monitoring, closed-loop theranostics, and smart drug delivery \cite{kuscu2021internet}. Beyond medicine, IoBNT applications extend to environmental monitoring, where networks of BNTs could collaboratively detect toxins or pathogens in ecosystems and supply chains, and smart agriculture, enabling molecular-level monitoring of crop health for optimized interventions \cite{babar2024sustainable}. 

The IoBNT architecture comprises three fundamental components: (i) \textbf{BNTs} are the functional nodes, encompassing engineered nanodevices and modified biological cells. They perform tasks analogous to IoT devices, including sensing, signal processing, actuation, and communication at the molecular level. As BNT capabilities dictate the overall system capacity, their effective design and integration are crucial for realizing the IoBNT vision \cite{kuscu2021internet}. (ii) \textbf{MC} is the main communication modality. Unlike electromagnetic communication, MC encodes information in the physicochemical properties of molecules (e.g., type, concentration, release timing) \cite{kuscu2019transmitter}. This nature-inspired approach offers inherent biocompatibility and energy efficiency by mirroring intercellular signaling. However, MC presents unique engineering challenges, including noise arising from stochastic molecular diffusion and ligand-receptor binding, significant channel memory, and high nonlinearity within dynamic physiological environments \cite{jamali2019channel}. (iii) \textbf{Bio-cyber interfaces} are the gateways bridging the biochemical domain of the nanonetwork and the electrical/electromagnetic domain of macroscale networks (e.g., the Internet) \cite{zafar2021systematic}. They transduce signals between these disparate domains, enabling external monitoring and remote control of BNTs. These interfaces may include implantable, wearable, or on-skin devices that detect molecular signals (e.g., via biosensors) or transmit commands using physical stimuli (e.g., light, magnetic fields, acoustics) \cite{civas2021universal}.

\subsection{BNT Architectures and Limitations}
Over the past decade, research toward realizing IoBNT has led to several conceptual frameworks for BNTs. Existing implementations generally fall into three paradigms: (i) \textbf{nanomaterial-based devices}, (ii) \textbf{passive molecular agents}, and (iii) \textbf{biosynthetic devices}. While each offers unique capabilities, they also possess inherent limitations. These limitations motivate the proposal of a fourth paradigm explored in this paper: NG-CSE-enabled living cell BNTs. We now outline the existing paradigms, highlighting their characteristics and the necessity for the NG-CSE approach as a more programmable and biocompatible alternative.

\subsubsection{\textbf{Nanomaterial-Based Abiotic Devices - The Synthetic Approach}}
The first paradigm is defined by entirely synthetic, abiotic systems constructed from engineered nanomaterials. These BNTs aim to replicate the functions of electronic devices at the micro/nanoscale, integrating sensing, processing, and actuation without relying on biological components. This category includes diverse technologies. Graphene and carbon nanotubes (CNTs), owing to their exceptional physical and electrical properties, are prominent materials for highly sensitive biosensors \cite{civas2023graphene}. For example, Carbon Nanotube Field-Effect Transistors (CNT-FETs) enable real-time, label-free detection of single biomolecules by measuring changes in electrical conductance \cite{schroeder2018carbon}. This paradigm also includes active micro- and nanorobots for intrabody tasks like targeted drug delivery \cite{erkoc2019mobile}. Examples include Janus nanomotors, which self-propel via catalytic reactions (e.g., H${2}O{2}$ decomposition) \cite{ma2017bio}, and magnetically actuated nanoswimmers, which can be guided through the vasculature using external magnetic fields \cite{ceylan20193d}.

Despite their functional sophistication, nanomaterial-based BNTs face critical challenges for \emph{in vivo} deployed: (i) \emph{Biocompatibility and toxicity}: As foreign materials, their physicochemical properties (size, charge, surface chemistry) can elicit adverse biological responses, including oxidative stress and inflammation. Long-term bioaccumulation and cytotoxicity remain significant concerns. (ii) \emph{Biofouling}: In physiological fluids, nanomaterial-based device surfaces are rapidly coated by biomolecules. This biofouling can passivate the device, obstructing sensing elements or impeding motion. (iii) \emph{Energy dependency}: Lacking intrinsic metabolism, these abiotic devices require external energy, such as chemical fuels (e.g., potentially toxic levels of hydrogen peroxide) or externally applied physical fields (magnetic, acoustic, optical). This reliance complicates deployment in deep tissue or long-term \emph{in vivo} scenarios. (iv) \emph{Biological integration}: These purely abiotic BNTs lack the dynamic responsiveness and adaptability of living cells within native biological networks.

\subsubsection{\textbf{Passive Molecular Agents - The Emergent Network Approach}}
The second paradigm involves minimalist, single- or few-molecule agents. While individually simple, these agents can form emergent communication networks capable of transmitting information and demonstrating swarm-like capabilities. This category includes entities such as fluorescent proteins, quantum dots, and other nanoparticles, characterized not by complex internal machinery but by their collective ability to relay signals through fundamental physicochemical or biological phenomena. These agents function as network nodes where information transfer is an intrinsic, though often uncontrolled, capability.

Passive BNTs can exploit biological processes for communication, as demonstrated by the use of the coagulation cascade for amplified tumor targeting \cite{von2011nanoparticles}. In this system, transmitter nanoparticles localize at the tumor and activate coagulation, broadcasting the location via a fibrin clot. This clot is recognized by circulating receiver nanoparticles carrying a payload. Here, the coagulation cascade serves as the communication channel, significantly amplifying delivery. More complex, interactive communication loops have also been demonstrated \cite{llopis2017interactive}. For instance, a system using two populations of enzyme-powered nanoparticles established feedback: a sender population processed an input (lactose) to produce a messenger (glucose), activating a receiver population to release a second messenger (N-acetyl-L-cysteine). This second messenger provided feedback to the sender, triggering payload release. Beyond biological processes, physicochemical phenomena like Förster Resonance Energy Transfer (FRET) enable short-range, high-speed communication via non-radiative energy transfer between fluorophores \cite{kuscu2015fluorescent}.

Despite these demonstrations, passive BNTs are fundamentally limited by a lack of programmability and active control. The network topology is emergent and stochastic, governed by initial conditions (e.g., concentration and distribution) rather than active, logic-gated decisions. While information flow can be initiated by external or biochemical triggers, its propagation remains passive. This inherent lack of autonomy and programmability renders these networks unsuitable for advanced IoBNT applications requiring dynamic, intelligent, and controllable responses.

\subsubsection{\textbf{Biosynthetic Devices - The Genetic Engineering Approach}}

The third paradigm utilizes genetically engineered living cells as BNTs. This approach harnesses the inherent capabilities of biological systems by programming cells (bacterial or mammalian) with synthetic genetic circuits or metabolic pathways. Examples include bacteria engineered to detect and produce signaling molecules, or human cells reprogrammed for \emph{in vivo} therapeutic tasks \cite{zhu2024synthetic, sezgen2021multiscale}. These living devices leverage the advantages of biological complexity, such as mobility, adaptation, and self-regulation, and utilize native mechanisms (e.g., quorum sensing, gene regulation) for more complex functions. 

Despite their potential, biosynthetic BNTs present significant challenges. Genetic modification inherently raises biosafety concerns and faces stringent regulatory oversight. Risks include unintended off-target genetic effects, horizontal gene transfer (e.g., antibiotic resistance) to native microbes, and adverse immune responses \cite{zalatan2024engineering}. Moreover, synthetic gene circuits impose a substantial metabolic burden on the host cell, consuming resources essential for survival \cite{wu2016metabolic}. This burden can impair viability, lead to the silencing or mutation of the genetic circuit, and cause unpredictable behavior, particularly in complex \emph{in vivo} environments.

\subsubsection*{\textbf{Discussion}} 
A comparison of these three paradigms reveals a critical gap in the BNT design space. The progression from abiotic to biosynthetic BNTs highlights a persistent trade-off between functional complexity and the risks of biological integration. Nanomaterial-based BNTs offer sophistication but carry high biocompatibility risks. Passive BNTs minimize this risk but sacrifice functional complexity and autonomy. Biosynthetic BNTs leverage living platforms for high functionality, but genetic modification introduces severe risks concerning genetic stability, safety, and ethics. Consequently, no existing paradigm successfully merges programmability, biocompatibility, and autonomous responsiveness without the liabilities of toxicity or genetic alteration. This gap motivates a fourth paradigm: an approach that endows living cells with programmable functions without altering their genetic blueprint.

\subsubsection{\textbf{The Fourth Paradigm: Surface-Engineered Living Cell BNTs}}

NG-CSE is an emerging bioengineering approach that has the potential to create programmable, living BNTs by modifying the cell surface of natural living cells without altering their genes. In this paradigm, a living cell is transiently converted into a functional device through the external attachment of synthetic molecular components to its membrane. A variety of biochemical methods exist to achieve this, including hydrophobic insertion of lipid-anchored molecules, covalent conjugation of polymers or nanoparticles, enzyme-mediated attachment of functional groups, or use of high-affinity aptamers that bind to specific cell-surface targets. These exogenous surface elements act as modular interfaces, serving as sensors, signal transducers, logic gates, or actuators on the cell's exterior.
\begin{figure*}[t]
    \centering
    \includegraphics[width=\textwidth]{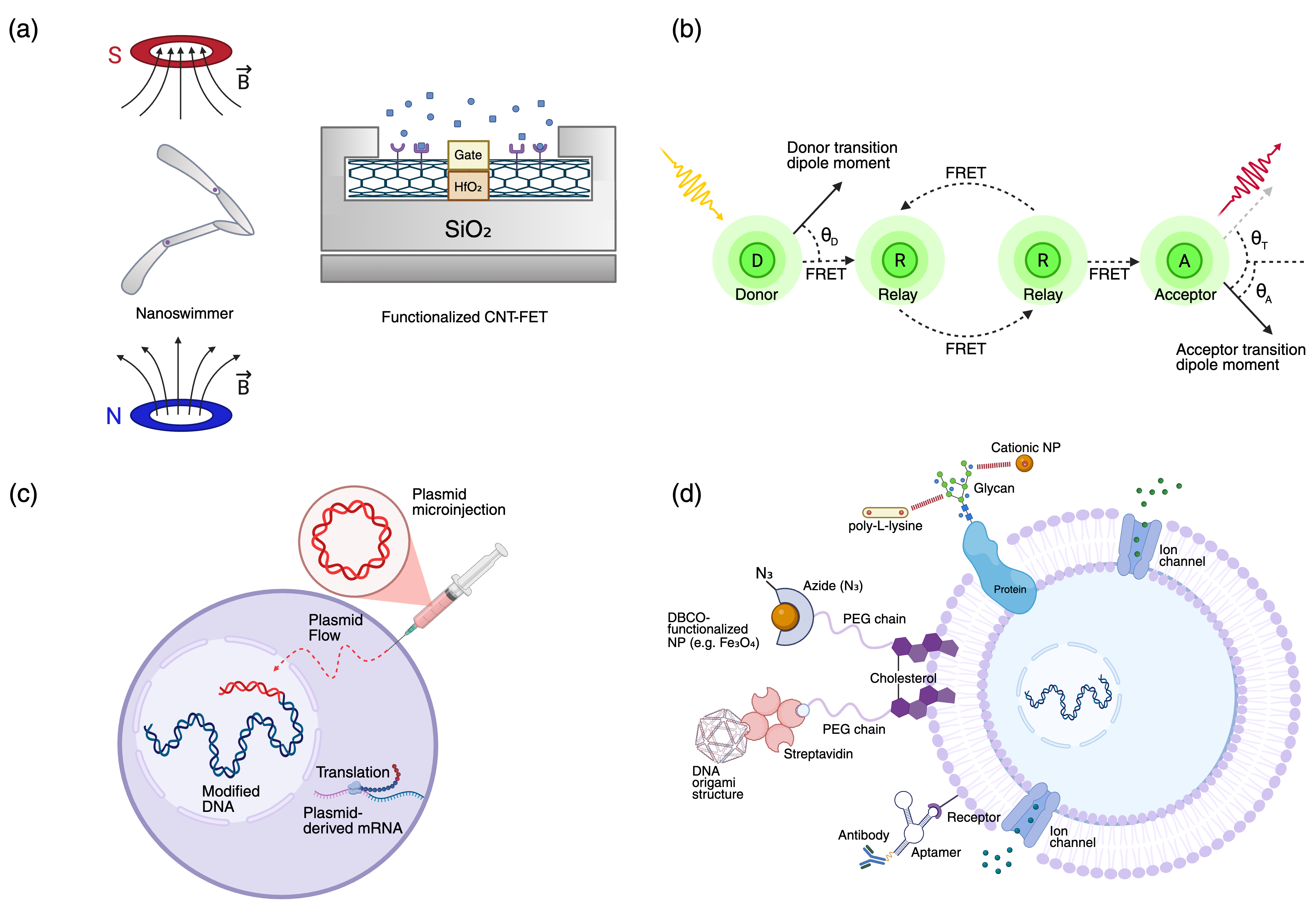}
    \caption{The landscape of BNT architectures for IoBNT. Comparison of the four main BNT architectures. (a) Nanomaterial-based abiotic devices offer high programmability but suffer from toxicity and lack agency. (b) Passive molecular agents are biocompatible but lack programmability and autonomy. (c) Biosynthetic devices (genetically engineered cells) offer agency but raise biosafety concerns and impose metabolic burdens. (d) The fourth paradigm, NG-CSE-enabled living cell BNTs, merges cellular agency with synthetic programmability without genomic modification. (Created with BioRender.com)}
    \label{fig:BNT}
\end{figure*}
Importantly, because NG-CSE avoids genomic modification, the engineered cell retains its native metabolism and viability, circumventing the safety concerns, metabolic burden, and unpredictability associated with genetic engineering. Concurrently, the cell acquires dynamically programmable features, functioning as a reconfigurable device capable of on-demand communication with other BNTs or electronic systems. This design approach synergistically integrates the adaptive intelligence and agency of autonomous living cells with the precise control afforded by nanotechnology. NG-CSE can thus enable a new class of intelligent, inherently biocompatible BNTs capable of participating in MC networks, responding to real-time stimuli, and interfacing with electronic infrastructure. The remainder of this paper investigates NG-CSE as a powerful strategy for constructing living, programmable BNTs that can overcome the limitations of existing BNT architectures.

\begin{figure*}[t]
    \centering
    \includegraphics[width=1\linewidth]{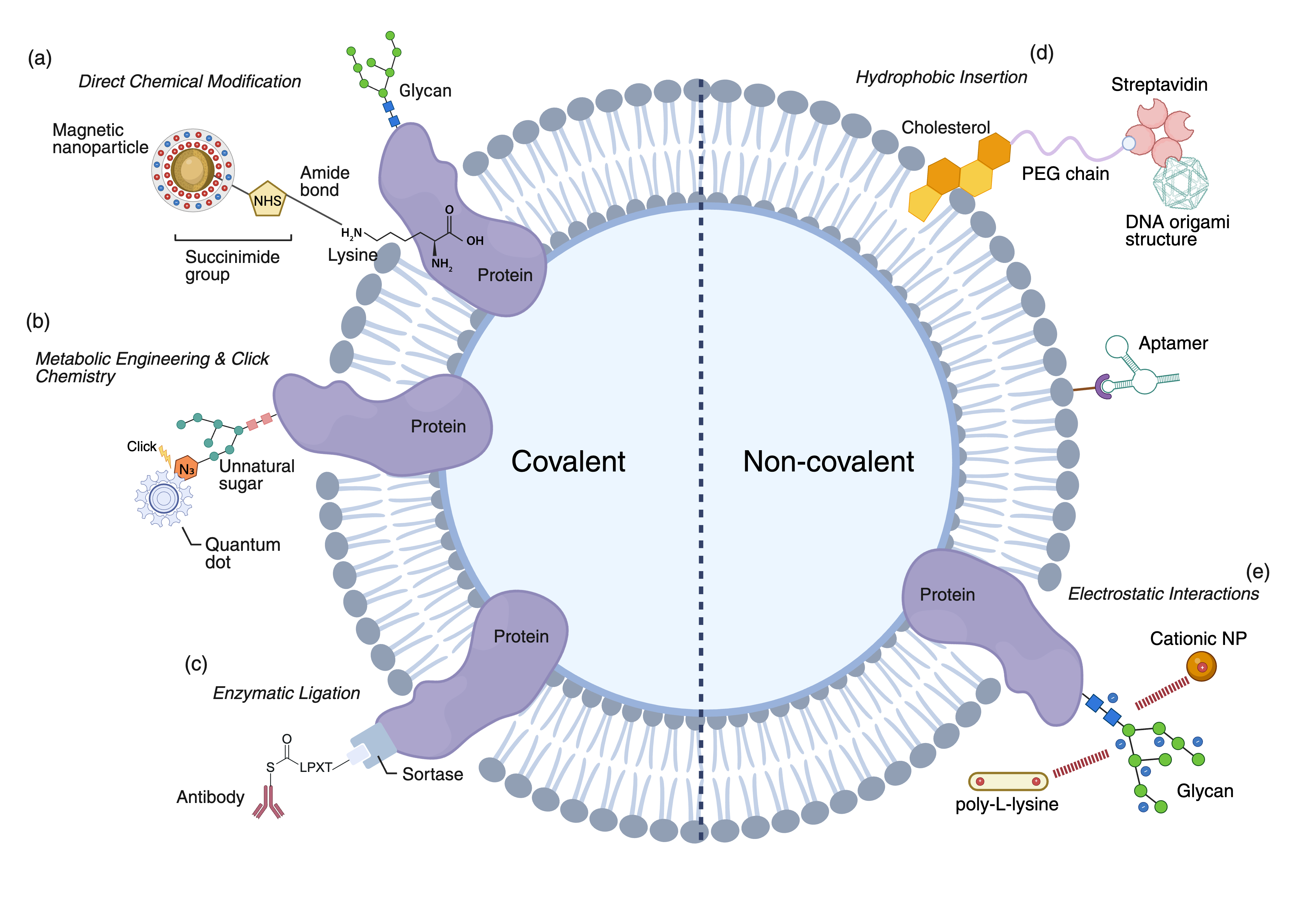}
    \caption{The toolbox for Non-Genetic Cell Surface Engineering (NG-CSE). NG-CSE employs diverse methods and materials to transform the cell membrane into a programmable interface. Engineering Methods are categorized by linkage type. Covalent strategies provide robust anchoring via (a) Direct chemical modification of native proteins, (b) Metabolic engineering and subsequent bioorthogonal click chemistry, or (c) Site-specific enzymatic ligation. Non-covalent strategies rely on physical interactions, including (d) Hydrophobic insertion of lipid- or cholesterol-anchored materials, and (e) Electrostatic interactions. Functional Materials impart new capabilities, including DNA nanostructures for programmable sensing and logic, nanoparticles for actuation and imaging, polymers (e.g., PEG) for immuno-cloaking, and proteins (e.g., antibodies) for targeting. (Created with BioRender.com)}
    \label{fig:toolboxandmethods}
\end{figure*}

\section{Hijacking the Membrane: The Interface and Toolbox}
\subsection{The Cell Membrane as a Programmable Interface in IoBNT}

The cell membrane is not merely a passive barrier but a dynamic, semipermeable interface that governs all interactions between a cell and its extracellular environment. This complex and fluid mosaic, composed of a phospholipid bilayer decorated with proteins and glycans, mediates a vast array of critical biological processes, including signal transduction, molecular transport, and cell-cell adhesion \cite{wu2024cell}. Because the cell surface serves as the main gateway for communication and interaction, it represents a critical target for therapeutic intervention and bioengineering. By precisely modifying or functionalizing this surface, it is possible to directly influence cellular physiology, control cell-microenvironment interactions, and program novel biological behaviors without breaching the intracellular space. 

This capability has positioned NG-CSE at the forefront of modern biotechnology with applications spanning critical areas of healthcare field. For example, in cancer immunotherapy, the surfaces of immune cells have been re-engineered to enhance their ability to recognize and destroy tumor cells, or whole cancer cells have been modified to serve as potent vaccines \cite{ma2021functional}. In regenerative medicine, stem cell surfaces are being functionalized to improve their survival, homing to injury sites, and to guide their differentiation into desired tissues \cite{park2018engineering}. Moreover, Red Blood Cells (RBCs, erythrocytes) have been repurposed as long-circulating drug delivery vehicles, carrying therapeutic payloads to specific targets within the body \cite{sun2017surface}. NG-CSE is not limited to mammalian cells such that bacterial cell surfaces have also been modified for medical applications. For example, in cancer therapy, attenuated bacteria, such as Salmonella, coated with nanoparticles, are being explored as tumor-targeting agents, as they naturally colonize hypoxic tumor cores \cite{suh2019nanoscale}. 

On the same grounds and motivated by the existing and emerging applications of NG-CSE, the cell membrane represents a highly strategic access point for IoBNT, offering structural, chemical, and functional anchoring grounds for external engineering of the cells to transform them into BNTs without the need for genetic modification. In the sequel, we will provide an overview of diverse materials and engineering approaches that enable cell surface modifications.

\subsection{Toolbox and Methods for NG-CSE)}
Hijacking living cells without altering their genome requires precise, biocompatible strategies to modify their surfaces. The success of this approach relies on two complementary domains: (1) functional materials that integrate with the membrane, and (2) engineering methods that ensure stable and selective attachment. This section outlines the key components of this toolbox (Fig. \ref{fig:toolboxandmethods}).

\subsubsection{\textbf{Functional Materials for Membrane Hijacking}}

The field has developed a vast and diverse toolkit of materials, ranging from synthetic polymers and lipids to inorganic nanoparticles, DNA nanostructures, and other functional biomolecules. These materials are also often combined in modular designs to integrate multiple functions into cell membranes.

\paragraph{\textbf{DNA Nanostructures}}  
DNA has emerged as an exceptionally versatile and programmable biomaterial within the NG-CSE toolbox. Single-stranded DNA aptamers can be selected and engineered to bind with remarkable specificity and high affinity to particular membrane proteins, functioning effectively as targeting ligands or as modulators of receptor-mediated cellular responses \cite{altman2013modifying, ueki2017nongenetic}. More complex, self-assembled DNA origami structures, created by folding a long scaffold strand using hundreds of short staple strands, can serve as rigid nanoscale breadboards with unparalleled spatial control, allowing other functional molecules to be positioned with nanometer precision on the cell surface, for example, to control receptor clustering \cite{nan2025advancements}. These structures can also function as mechanical actuators, protective shells and logic gates capable of responding intelligently to multiple biological or environmental stimuli \cite{feng2021recent, guo2022manipulation}.

\paragraph{Nanoparticles} Nanoscale materials (e.g. inorganic nanoparticles, quantum dots) can be attached to cells to impart new capabilities in imaging, sensing, or drug delivery \cite{grupi2019interfacing}. For example, by decorating cells with superparamagnetic iron oxide (Fe$_3$O$_4$) nanoparticles, cells can be manipulated or localized using external magnetic fields \cite{dobson2008remote}. In \cite{zhao2021surface}, natural killer immune cells (NK-92 line) were labeled with anti-CD56 antibody-conjugated Fe$_3$O$_4$ nanoparticles, which bound specifically to the NK cells' surface and endowed them with magnetic responsiveness. Nanoparticles can also turn cells into carriers for therapeutic payloads: scientists have attached drug-loaded nanoparticles to circulating cells so that the cells ferry these hitchhiking drugs to disease sites \cite{wang2024cell, stephan2011enhancing}. Other uses of cell-bound nanoparticles include imaging and biosensing: quantum dots or silica nanoparticles on cell surfaces can report on cell location or local biochemical signals \cite{lee2023nanoparticles, ali2012cell}.

\paragraph{Polymers} Natural or synthetic polymers are frequently used to coat or shield cells and to introduce new functionality \cite{Ghezzi2021-hd}. One common polymer is poly(ethylene glycol) (PEG), which can be covalently grafted to cell membranes to camouflage cells from the immune system, a process analogous to PEGylation of drugs \cite{wang2015non}. For example, RBCs coated with PEG or hyperbranched polyglycerol have reduced immunogenicity and avoid rapid clearance, making transfusions more compatible between donors and recipients. Cationic polymers like poly-L-lysine (PLL) or poly(ethyleneimine) (PEI) are used in layer-by-layer encapsulation to form polymeric shells around cells \cite{lee2018cell}. These nanoscale coatings can protect cells from hostile environments and mechanically confine secreted factors for tissue engineering applications. Advanced polymer chemistry has also enabled in situ formation of polymer networks on cell surfaces. For example, researchers have inserted polymerization initiators into cell membranes and then carried out controlled radical polymerization, growing polymer brushes directly from the cell surface \cite{niu2017engineering}. Such approaches allow creation of a cell-bound hydrogel layer or synthetic glycocalyx that can modulate cell–cell interactions and signaling \cite{zhong2023site}.

\paragraph{Proteins and Peptides} Grafting proteins onto a cell surface can bestow the cell with new biological functions or targeting abilities \cite{henry2020kodecytes}. Protein-based surface modifications are powerful as they leverage highly specific biological interactions, e.g., antigen-antibody, ligand-receptor. A common strategy is attaching antibodies or fragments thereof to cell membranes to target the cells to specific tissues \cite{ma2023advances}. For example, coating mesenchymal stem cells (MSCs) with a polyvalent display of antibodies have been shown to greatly improve their homing to injured tissue \cite{ye2023nongenetic}. Likewise, peptide ligands that mimic selectin ligands or contain integrin-binding motifs have been grafted onto the cell surfaces of MSCs, significantly improving their adhesion to target endothelium and trafficking to injury sites \cite{park2024ex}. Moreover, in \cite{wu2023chemically}, chemically synthesized, modular synthetic receptor proteins have been demonstrated to detect acidic pH, a hallmark of tumor microenvironments, and trigger specific intracellular signaling pathways on demand, providing new sense-and-respond capabilities imparted entirely by NG-CSE.

\subsubsection{\textbf{Cell Surface Engineering Methods}}

The successful deployment of functional materials onto the cell membrane to create living BNTs can be realized with a diverse array of chemical and physical strategies. Broadly, these methods can be categorized based on the nature of the linkage: covalent conjugation, which forms relatively stable chemical bonds, and non-covalent conjugation, which relies on physical interactions \cite{almeida2023cell, roy2020nongenetic}. The choice between these methods determines the level of stability, specificity, and reversibility of the modification, directly impacting the functional lifecycle of the resulting BNT.

\paragraph{Covalent Conjugation} 
Covalent strategies enable relatively robust linkages between synthetic materials and the cell surface. This stability is particularly desirable for IoBNT applications requiring long-term deployment or resilience in dynamic physiological environments. Covalent modification can be achieved through direct chemical reactions with native membrane components, or indirectly via metabolic or enzymatic engineering to introduce specific chemical handles.

\begin{itemize} 
\item \textbf{Direct Chemical Modification:} This method targets naturally occurring nucleophilic groups on the cell surface, primarily the amines (e.g., lysine residues) and thiols (e.g., cysteine residues) of membrane proteins \cite{csizmar2018programming, liu2019advances}. Common strategies include the use of N-hydroxysuccinimide (NHS) esters for amine coupling and maleimides for thiol coupling \cite{abbina2017surface}. While providing rapid functionalization, the method lacks specificity, potentially modifying any accessible surface protein. This indiscriminate labeling can disrupt native protein function and cellular signaling \cite{almeida2023cell}, potentially compromising the agency of the BNT. However, for applications where robust anchoring of materials (e.g., polymeric shields or sensor arrays) is prioritized over precise spatial control, direct chemical modification remains a viable strategy.

\item \textbf{Metabolic Engineering:} To achieve greater control and biocompatibility, metabolic engineering exploits the cell's biosynthetic machinery to introduce bioorthogonal chemical reporters onto the cell surface \cite{roy2020nongenetic}. This is frequently achieved through metabolic glycoengineering (MGE), where cells are fed unnatural sugar analogs (e.g., bearing azide or ketone groups) that are processed and displayed within the cell's glycocalyx \cite{almeida2023cell, csizmar2018programming}. These unique chemical handles can then be addressed with high specificity using bioorthogonal chemistries, such as the Staudinger ligation or click chemistry, to attach functional materials \cite{roy2020nongenetic}. MGE allows for the installation of programmable interfaces without direct chemical intervention in the membrane. For IoBNT, this enables the precise attachment of functional components while minimizing perturbation to the underlying cellular machinery. The modifications are relatively transient, dictated by the natural turnover of the glycans, facilitating the \emph{reversible} hijacking of the cell \cite{almeida2023cell}.

\item \textbf{Enzymatic Ligation:} Enzymatic methods offer high selectivity under mild physiological conditions by using enzymes to catalyze the formation of covalent bonds at specific sites \cite{almeida2023cell}. Enzymes such as Sortase A or lipoic acid ligase can be used to recognize specific peptide sequences (which may be native or introduced as fusion tags) and ligate them to functionalized substrates \cite{csizmar2018programming, abbina2017surface}. Moreover, enzymes like galactose oxidase or glycosyltransferases can remodel existing glycans to introduce or modify functional groups \cite{almeida2023cell}. Enzymatic ligation provides exquisite site-specificity, which is crucial for the rational design of BNTs where the precise localization of synthetic machinery (e.g., positioning a sensor component relative to a native receptor) is critical for function.

\end{itemize}

\paragraph{Non-Covalent Conjugation}
Non-covalent strategies rely on physical interactions or affinity binding to anchor materials to the cell surface. These methods are often characterized by their simplicity, rapid execution, and high biocompatibility, as they typically do not alter the chemical structure of native membrane components. They are particularly advantageous for IoBNT scenarios requiring more rapid deployment, transient functionalization, or the stronger preservation of native cellular behavior.

\begin{itemize} 
\item \textbf{Hydrophobic Insertion:} This widely used technique exploits the amphipathic nature of the plasma membrane. Functional materials conjugated to hydrophobic moieties, such as cholesterol, phospholipids, or alkyl chains, spontaneously insert into the lipid bilayer \cite{liu2019advances, schoenit2021functionalization}. This method is rapid, \emph{universal} (as it does not depend on specific protein expression), and generally preserves cell viability \cite{csizmar2018programming}. DNA nanostructures, for example, are frequently anchored using cholesterol tags \cite{schoenit2021functionalization, jia2022recent}. However, the modifications are inherently dynamic; the anchored materials can diffuse laterally within the membrane and may be internalized or dissociate over time \cite{almeida2023cell}. For IoBNT, hydrophobic insertion is ideal for transiently programming cell interactions. 

\item \textbf{Electrostatic Interactions:} The net negative charge of the mammalian cell surface, conferred by sialic acid residues and phosphate groups, can be utilized to adsorb positively charged materials, such as cationic polymers (e.g., poly-L-lysine) or nanoparticles \cite{almeida2023cell, abbina2017surface}. This is often employed in the layer-by-layer (LbL) technique to construct thin polyelectrolyte films around the cell \cite{liu2019advances}. While simple, this approach is non-specific and high densities of positive charge can disrupt membrane integrity and induce cytotoxicity \cite{almeida2023cell}. It is most suitable for applications focused on encapsulation or shielding the BNT, e.g., for immune escape, rather than precise functionalization.

\end{itemize}

\section{Opportunities for IoBNT enabled by Surface-Engineered Living Cells}

Surface-engineered cells are promising candidates for living BNTs in IoBNT applications, as they can be transiently programmed with new sensing and communication interfaces, and integrate other key functionalities that meet the demands of various IoBNT applications, without needing genetic modification. Here, we discuss the critical capabilities that NG-CSE unlocks for living BNTs.

\subsection{Reprogramming Cellular Interactions}

Since the plasma membrane is the interface between a cell and its environment, non-genetic modification of this surface provides direct, on‑demand control over the cell's input–output characteristics, enabling several key opportunities for IoBNT integration.

\subsubsection{\textbf{Creation of De Novo Sensing Functions}} NG-CSE can expand a cell’s sensory repertoire beyond its native capabilities. By tethering artificial receptors, aptamers, or functional nanomaterials to the membrane, cells can be rendered responsive to abiotic or synthetic molecular cues, such as those emitted by nanomaterial-based BNTs. For example, a DNA-mediated chemically induced dimerization (D-CID) nanodevice has been developed to reprogram endogenous receptors \cite{li2018dna}. This system uses DNA aptamers or DNAzymes as recognition modules that, upon binding abiotic small-molecule cues (e.g., Zn$^{2+}$), trigger a toehold-mediated strand displacement cascade. This cascade induces the dimerization and activation of the endogenous c-Met receptor, conferring new responsiveness to these inputs and controlling downstream functions like cell growth. A complementary approach demonstrated the integration of a fully synthetic receptor into the membrane, establishing an artificial sense-and-respond signaling pathway \cite{wu2023chemically}. Such modular receptors establish direct communication channels between artificial and biological nodes. This enables sophisticated IoBNT tasks, including stimulus-gated drug release or chemotactic navigation, triggered by unique, addressable molecular keys, which is conceptually analogous to assigning an Internet Protocol (IP) address to a BNT.

\begin{figure*}[t]
    \centering
    \includegraphics[width=1\linewidth]{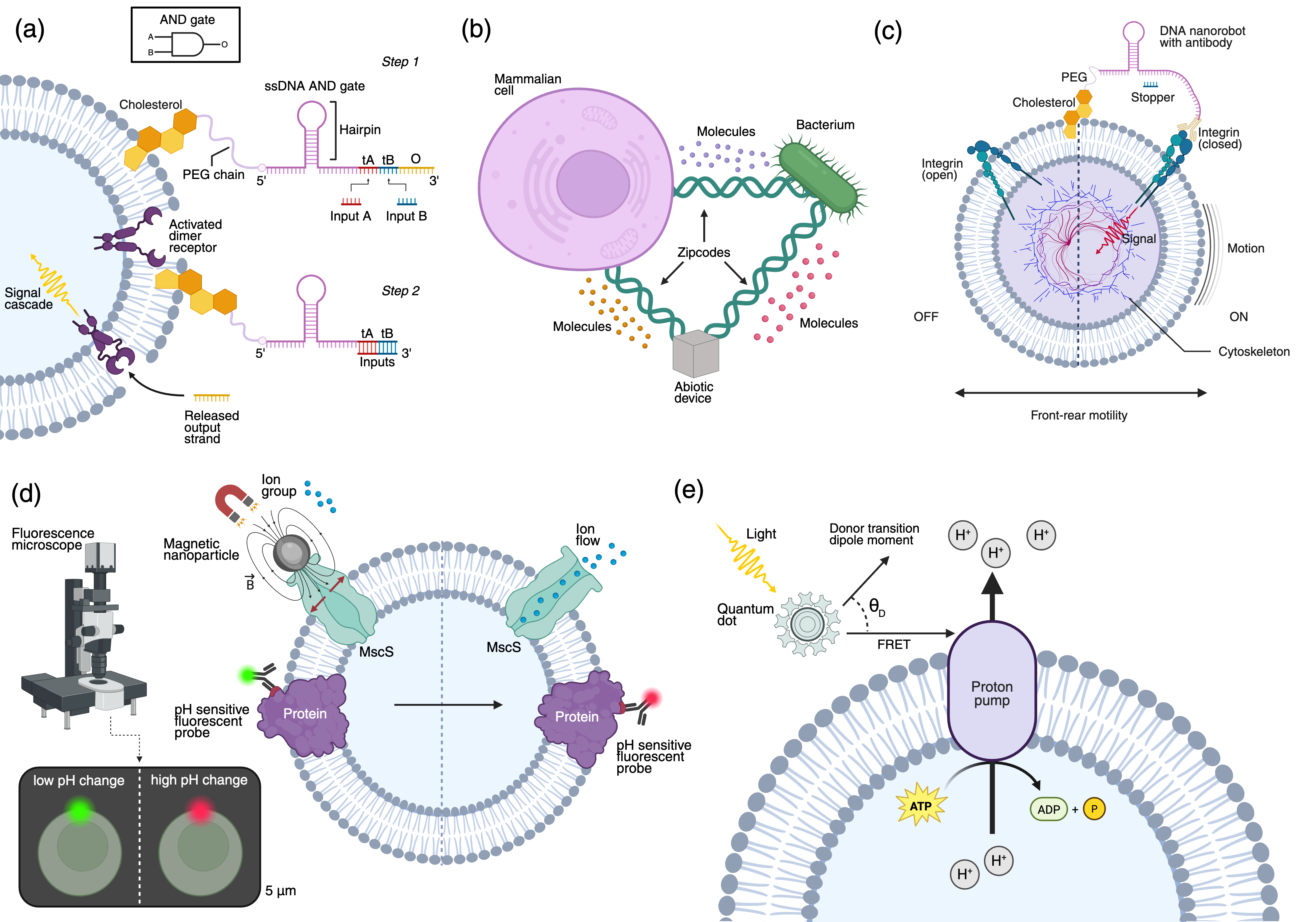}
    \caption{Key opportunities for IoBNT enabled by NG-CSE. NG-CSE transforms living cells into dynamic, programmable BNTs. (a) Cell-Surface Logic and Sensing. Membrane-anchored DNA logic gates (e.g., AND gate) process multiple molecular inputs to execute localized computation and control endogenous receptor signaling. (b) Dynamic Network Topologies. Programmable adhesion molecules facilitate the rational assembly of heterogeneous networks, enabling interkingdom communication and integration with abiotic devices. (c) Modulation of Mobility. Cellular agency can be externally controlled. Reconfigurable DNA nanorobots act as molecular switches to reversibly activate (ON) or halt (OFF) cell migration by controlling integrins. (d) Bio-Cyber Interfacing. Surface-tethered nanoparticles enable remote actuation via external fields (e.g., magnetic input), while optical probes provide real-time environmental readout (output). (e) Energy Harvesting. The cellular energy budget can be augmented by installing surface machinery, such as Quantum Dot-protein hybrids, that harvest light energy (e.g., via FRET) to enhance metabolic ATP production. (Created with BioRender.com)}
    \label{fig:placeholder}
\end{figure*}
\subsubsection{\textbf{Cell-Surface Logic and Multi-Signal Processing}} The concurrent installation of multiple orthogonal surface receptors can enable living BNTs to execute Boolean logic operations directly at the membrane interface. Proof-of-concept studies have demonstrated logic gating using membrane-anchored DNA nanodevices that process multiple inputs to modulate endogenous receptor signaling. For example, YES, NOR, AND, and OR gates have been constructed using c-Met and CD71 aptamers to control receptor assembly, thereby modulating the HGF/c-Met signaling pathway and regulating cell migration \cite{chen2019logic}. In another demonstration, an AND logic nanodevice was engineered to respond to the simultaneous presence of protons (H$^{+}$) and ATP, key biomarkers of the tumor microenvironment. This device selectively assembled MET and CD71 receptors only when both inputs were present, leading to the inhibition of cancer cell motility \cite{wu2023logic}. In an IoBNT context, these capabilities allow a living BNT to execute a physiological function, such as drug payload release, only when it receives an external command from the network AND detects a specific combination of local biomarkers. 

\subsubsection{\textbf{Dynamic Network Topologies and Distributed Systems}} Biological functionality often emerges from the orchestrated interactions of distinct cell types. NG-CSE permits the \emph{in situ} and programmable assembly of cells into controlled, dynamic, and hierarchically structured topologies, a key enabler for IoBNT. By installing complementary adhesion or signaling molecules on the membranes of selected populations, their autonomous self-assembly into synthetic consortia with emergent behaviors exceeding individual capabilities can be directed. For example, a membrane-anchored DNA nanoplatform enabled the precise, tunable, and reversible assembly of cells into clusters, with interaction strength controlled by the linking DNA sequence length \cite{li2019cell}. Furthermore, DNA origami nanostructures have been utilized as biomimetic membrane channels to organize cells into \emph{cell origami clusters} (COCs) with programmable topologies (e.g., linear vs. closed-ring) \cite{ge2020programming}. More importantly, this method can establish links between otherwise incompatible biological systems, enabling \emph{inter-kingdom communications} \cite{wang2023dna}. The resulting ability to fabricate rationally designed multicellular networks can facilitate IoBNT architectures where bacteria, immune cells, and synthetic microsensors operate cooperatively. Programmable control over nodal interactions and information transmission allows for the elicitation of novel collective behaviors and, indeed, an entirely new \emph{collective physiology}. Such distributed, heterogeneous networks with on-demand connectivity can enable sophisticated sensing-and-response loops, leveraging the unique strengths of each cell type for optimized IoBNT performance.

\subsubsection{\textbf{Improved IoBNT Persistence and Niche Targeting}} The long-term efficacy of IoBNT applications depends on the capacity of BNTs to persist and operate within dynamic and often immunologically hostile physiological environments. NG-CSE provides a versatile toolkit for enhancing the resilience of living BNTs. Covalent grafting of stealth polymers, such as PEG, onto cellular surfaces, a strategy termed \emph{immuno-cloaking}, allows BNTs to evade immune surveillance, prolonging their functional lifetime in vivo \cite{kim2018general, kim2015recent}. 

Beyond immune evasion, surface engineering facilitates the targeted deployment of BNTs to otherwise inaccessible locations. Functionalizing a BNT's surface with synthetic adhesion molecules enables anchoring and operation in specific niches where they would naturally be cleared. For example, a bacterial BNT could be surface-modified to adhere to the tumor microenvironment, establishing a localized, drug-producing biofilm or functioning as a persistent biosensor, thereby serving as a stable node within the IoBNT network \cite{fu2024surface}. NG-CSE also affords a means to insulate engineered communication pathways from biochemical noise in the host environment. Selective masking or chemical blockade of native receptors minimizes interference from endogenous ligands, significantly improving the signal-to-noise ratio of externally delivered IoBNT commands \cite{song2024engineering}. This targeted decoupling from natural inputs can ensure that the IoBNT application is governed by designed control signals rather than stochastic physiological fluctuations, enhancing reliability and predictability.

\subsection{Modulation of Mobility} 

Control over BNT motility is critical for many envisioned IoBNT applications. NG-CSE techniques provide a portfolio of orthogonal strategies, such as mechanical, chemical, and adhesive, that can (i) initiate motility in sessile cells, (ii) tune velocity over orders of magnitude, (iii) impose directional bias, and (iv) orchestrate emergent swarm-level patterns, all without altering the host cell's genome.

\subsubsection{\textbf{Activating Intrinsic Motility Pathways}} Synthetic receptor-clustering devices attached to the cell surface can function as reversible molecular switches for programming motility. This has been exemplified by aptamer-gated DNA nanorobots that dimerize the MET receptor \cite{li2023transformable}. In the closed conformation, the surface-bound nanorobot induces receptor dimerization, activating downstream signaling pathways that initiate cell migration. A DNA strand displacement reaction, triggered by an opener strand, shifts the nanorobot to an open state, separating the receptors and halting migration. This dynamic and reversible switch can, therefore, provide a high level of precise, real-time control over living cell BNT motility without genetic engineering.

\subsubsection{\textbf{Attenuating Cell Motility}} Conversely, NG-CSE can be used to restrain movement of living cell BNTs. A representative example is a stimuli-responsive DNA nanodevice that self-assembles into extended DNA duplexes on the plasma membrane in response to elevated ATP concentrations characteristic of tumor microenvironments \cite{su2022enhancing}. The resulting nucleic-acid scaffold improves lipid-raft clustering and membrane phase separation, generating a steric barrier that prevents adhesion receptors from engaging the extracellular matrix. Consequently, cancer-cell migration and metastasis are markedly suppressed both \emph{in vitro} and \emph{in vivo}. By imposing this reversible molecular brake, motile cells can be converted into site-anchored BNTs without compromising their native sensing or effector functions, expanding the design space for living cell BNTs.

\subsubsection{\textbf{Programming Collective Dynamics and Movement Patterns}} DNA nanostructures on the cell surface can be utilized not only to modulate single-cell behavior but also to re-engineer the emergent dynamics of cell collectives. Recent studies on epithelial monolayers have shown that the insertion of amphiphilic lipid-DNA micelles increases the coupling between individual cell shape and velocity, thereby enhancing directional coordination within the sheet \cite{wang2023collective}. In contrast, the introduction of mechanically rigid DNA-origami tiles raises the effective elastic modulus of the monolayer and dampens coordinated motion, leading to a global reduction in migration speed. These orthogonal design methods provide a tunable toolkit for engineering swarm-level behaviors, a critical capability for orchestrating distributed BNT ensembles within an IoBNT framework.

\subsection{Bio-Cyber Interfacing} 

Interfacing BNTs with external controllers and communication networks (the cyber domain) is fundamental to IoBNT applications \cite{zafar2021systematic}. This interface must enable both the monitoring of BNTs and the molecular messages they exchange, and the remote modulation of their functions. Living cells, however, present limited channels for such communication, particularly \emph{in vivo}, making their integration more challenging than nanomaterial-based counterparts. While genetic engineering can introduce pathways, such as fluorescent reporters or optogenetic actuators \cite{emiliani2022optogenetics}, these methods are slow, often single-use, and impose a significant metabolic burden on the host.

A main requirement for bio-cyber interfacing is the real-time readout of BNT states and their microenvironment. NG-CSE offers several strategies to enable living BNTs to broadcast this information. The most direct approach is to functionalize the plasma membrane with sensing moieties that optically report on physicochemical cues \cite{ali2012cell, lee2023nanoparticles}. For example, integrated functional fluorescent probes can report on biophysical properties such as lipid order, membrane viscosity, and mechanical stress using specialized molecules, such as solvatochromic dyes and \emph{flippers} \cite{collot2022advanced}. Other probes can monitor local concentrations of specific analytes (e.g., pH, Zn$^{2+}$, metabolites) in real-time \cite{ali2012cell, collot2022advanced}. More sophisticated platforms utilize DNA-based nanosensors that transduce rapid, transient molecular encounters, such as microsecond-timescale lipid interactions, into a cumulative fluorescence signal via toehold-mediated strand displacement \cite{you2017dna}. Nanoparticles serve as versatile scaffolds for these sensing elements, providing higher brightness and photostability critical for long-term monitoring, overcoming limitations of small-molecule probes \cite{lee2023nanoparticles}.

IoBNT applications also require remote actuation of BNTs. NG-CSE provides a versatile toolbox for control via externally applied physical fields, mainly light and magnetism. MNPs attached to the cell surface can serve as remote actuators leveraging the excellent tissue penetration of magnetic fields for \emph{in vivo} applications \cite{dobson2008remote, zhang2023functional}. In magneto-mechanical stimulation, low-frequency magnetic fields induce MNP oscillation or rotation, exerting precise forces on the cell. High-gradient magnetic fields can apply translational pulling forces to activate mechanosensitive ion channels or induce receptor clustering \cite{dobson2008remote, nag2020nanoparticle}. Alternatively, magnetic twisting cytometry applies rotational torque to probe cytoskeletal mechanics or trigger downstream signaling \cite{dobson2008remote}. These techniques have been used to remotely induce neuronal firing by activating Piezo1 channels in freely moving animals \cite{zhang2023functional}, guide stem cell differentiation, and promote bone matrix mineralization \cite{dobson2008remote}.

Optical modalities provide complementary advantages, notably superior spatiotemporal precision for single-cell or region-specific addressing  \cite{huang2023nanoparticles}. In photothermal actuation, membrane-tethered plasmonic nanoparticles (e.g., gold nanorods (AuNRs)) absorb near-infrared (NIR) light and convert it into localized heat. This nanoscale temperature increase can directly activate thermosensitive ion channels (e.g., TRPV1) to drive calcium influx and electrical responses \cite{huang2023nanoparticles}, or trigger molecular cargo release. For example, the LSPR-based photothermal effect of AuNRs has been used to de-hybridize and release a DNA-based agonist, which subsequently dimerized native receptors to promote skeletal muscle regeneration \emph{in vivo} \cite{wang2019near}. Photomechanical actuation, a more advanced strategy, utilizes nanoparticles with a AuNR core and a thermo-responsive polymer shell. Upon NIR illumination, rapid shell collapse exerts piconewton-scale mechanical pulling forces on tethered receptors, enabling control over processes, such as focal adhesion formation and T-cell activation, with millisecond precision \cite{liu2016nanoscale}. Other methods include using upconversion nanoparticles to convert NIR to visible light for controlling light-gated channels in deep tissue \cite{huang2023nanoparticles}.

\subsection{Energy Harvesting}

Living cells operate on a tightly regulated energy budget, converting environmental chemical energy into biologically accessible forms to drive essential processes such as growth and homeostasis. While sufficient for native functions, this internal bioenergetic machinery may be insufficient for the demands of IoBNT applications, which impose energy-intensive tasks such as long-range motility, continuous multi-analyte sensing, and high-bandwidth molecular signaling. Augmenting the cellular power budget is, therefore, critical for deploying self-sustaining living BNTs. NG-CSE addresses this challenge by creating novel interfaces for energy augmentation \cite{grupi2019interfacing, hicks2021electric}.

One strategy is to install photonic harvesters on the cell surface, inducing artificial photosynthesis \cite{hicks2021electric}. This has been demonstrated with the non-photosynthetic bacterium Moorella thermoacetica, using inorganic nanoparticles as light-harvesting antennas. Upon illumination, these nanoparticles generate photo-excited electrons that are funneled into the bacterium's metabolic pathways, enabling the photosynthetic production of acetic acid from CO$_2$ \cite{hicks2021electric}. A similar principle involves coupling semiconductor QDs with bacteriorhodopsin (bR), a natural light-driven proton pump \cite{padma2023artificial}, \cite{rakovich2010resonance}. In this hybrid system, QDs act as nanoconverters, absorbing light in spectral regions where bR is inefficient (e.g., UV-blue) and transferring the energy to bR via FRET. This mechanism has been shown to significantly enhance bR's transmembrane proton pumping, which is a key step in ATP synthesis, by over 25\% \cite{padma2023artificial}.

A complementary strategy establishes a direct electrochemical interface with the cell, enabling on-demand modulation of redox-dependent energy pathways \cite{grupi2019interfacing, li2007stationary}. This has been realized using short, lipid-coated carbon nanotube porins (CNTPs), i.e., biomimetic CNTs that spontaneously insert into and span the cellular membrane \cite{li2007stationary}. These CNTPs function as wireless bipolar electrodes. When subjected to an external, low-voltage ($<$ 2 V) electric field, the nanotubes polarize and facilitate transmembrane electron transport without a physical connection \cite{li2007stationary}. This provides an on-demand, wireless method to control electron flux, potentially driving the reduction of intracellular electron acceptors and augmenting the cell's metabolic reducing power \cite{grupi2019interfacing,li2007stationary}.

\section{IoBNT Applications Enabled by NG-CSE}

The unique combination of cellular agency, transience, and programmability offered by NG-CSE enables novel IoBNT applications and architectures that are unattainable with other BNT paradigms. By rationally designing IoBNT systems that exploit the innate behaviors of living cells while programming their interactions through surface modification, high-impact IoBNT applications in diagnostics, unconventional computing, and therapeutics can be realized. Here, we discuss two such IoBNT applications that can be uniquely enabled by NG-CSE-enabled living BNTs.

\begin{figure*}[t]
    \centering
    \includegraphics[width=1\linewidth]{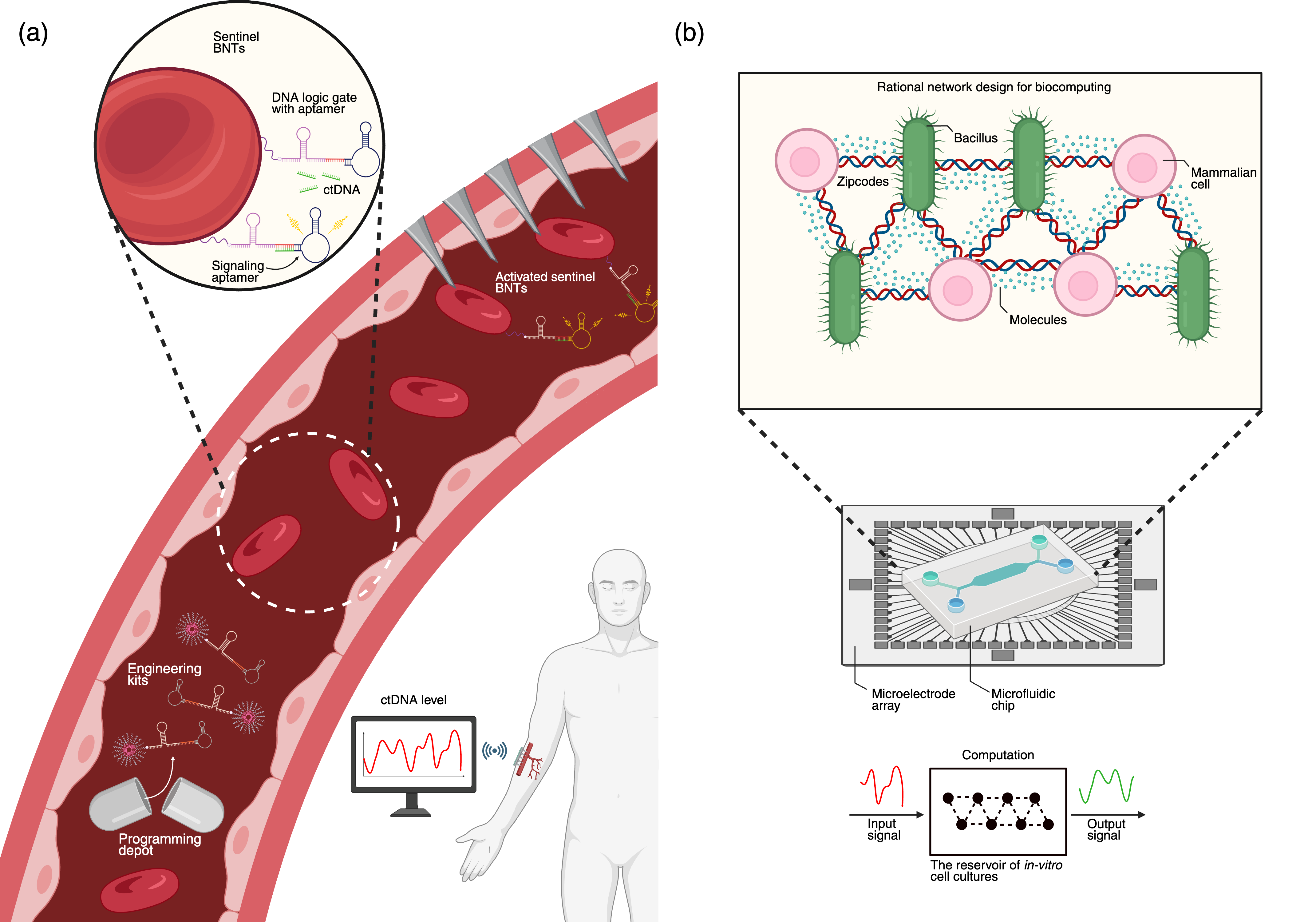}
    \caption{IoBNT Applications Enabled by NG-CSE. The combination of programmability and cellular agency enables novel IoBNT architectures. (a) Circulating Sentinel Network for Continuous Liquid Biopsy. (1) A programming depot releases molecular engineering kits. (2) These kits transiently hijack Red Blood Cells (RBCs) in situ, creating Sentinel BNTs. (3) Sentinels patrol the vasculature; a surface DNA logic gate detects ctDNA and activates a signaling aptamer. (4) A wearable microneedle patch (bio-cyber gateway) captures activated Sentinels and transduces the detection into an electronic signal. (b) Unconventional Computing with Rationally Designed Microphysiological Systems. (1) Heterogeneous cell populations (e.g., interkingdom consortium) are surface-engineered with DNA zip codes. (2-3) These codes direct the self-assembly of a structured network (biological reservoir) on a Micro-Electrode Array (MEA), contrasting with stochastic self-organization. (4) The MEA acts as the bio-cyber interface for input encoding and output readout, realizing a rationally designed "wetware" computer. (Created with BioRender.com)}
    \label{fig:placeholder}
\end{figure*}
\subsection{The Circulating Sentinel Network for Liquid Biopsy}
A major challenge in oncology is the need for continuous, body-wide surveillance for minute traces of circulating tumor DNA (ctDNA) \cite{gao2022circulating}. The merger of NG-CSE and IoBNT can uniquely enable a \emph{circulating sentinel network} for continuous liquid biopsy, i.e., body-wide surveillance for minute traces of circulating tumor DNA (ctDNA) by transiently hijacking RBCs. Mature RBCs are ideal surveillance agents due to their abundance, long lifespan, and natural circulatory agency; however, being anucleated, they cannot be genetically engineered. NG-CSE uniquely enables their functionalization \emph{in situ} \cite{brenner2021red}. The network is initiated by an implantable or injectable \emph{programming depot} (e.g., an abiotic BNT transmitter) that releases molecular engineering kits, such as aptamer-lipid conjugates. These conjugates spontaneously insert into the membranes of passing RBCs via hydrophobic insertion, converting them into sentinel BNTs. The surface of these sentinels is engineered with a DNA-based logic gate comprising a ctDNA-specific capture probe and a sequestered signaling aptamer. As the sentinel BNTs utilize their natural agency to patrol the vasculature, the binding of target ctDNA triggers the logic gate via strand displacement, unmasking the signaling aptamer. This logic-gated MC ensures that only positive sentinels are activated for readout. A wearable microneedle patch, acting as the bio-cyber gateway of the IoBNT, interfaces with the vasculature. The microneedles are functionalized to bind the unmasked signaling aptamer. The capture of Positive Sentinels at the microneedle interface is transduced into an electronic signal, bridging the biochemical detection event with the cyber domain for continuous oncology surveillance.

\subsection{Unconventional Computing with \emph{in vitro} Microphysiological Systems with Programmed Self-Assembly and Interactions}

A second compelling application domain is the development of rationally designed microphysiological systems to realize unconventional computing, exploiting the innate information processing capacity of biological systems. Current efforts in biological computing, such as Organoid Intelligence (OI) framework \cite{cai2023brain}, rely on the stochastic self-organization of biological matter, leading to high variability, making these substrates unreliable as a computational resources. NG-CSE can enable the rational design of the biological reservoir by precisely programming the adhesion, spatial organization, and communication topology of heterogeneous cell populations. In this IoBNT-on-a-chip architecture, distinct populations of living cells, potentially integrating human cells and specific bacterial strains to maximize dynamic complexity, are surface-engineered with unique, complementary DNA zip codes via lipid anchoring. When introduced into a microfluidic device integrated with a micro-electrode array (MEA), these programmed surface addresses dictate the self-assembly of a structured, reproducible, three-dimensional \emph{interkingdom network}. This rationally designed microphysiological system serves as the computational reservoir.  The MEA functions as the bio-cyber interface, encoding input data into spatiotemporal stimulation patterns and reading out the network's complex, high-dimensional response. By using NG-CSE to engineer the physical connectivity and exploit the novel dynamics of interkingdom signaling, this IoBNT approach can move beyond emergent computation to rationally designed biological computation, enabling reproducible, high-performance \emph{wetware} computers.

\section{Challenges and Future Research Directions}

The use of NG-CSE to create living BNTs could be a paradigm shift for the IoBNT. By leveraging the cell's innate viability while externally programming its surface, NG-CSE circumvents the risks of genetic modification and the biocompatibility issues of purely abiotic devices. However, the very features that make NG-CSE unique, particularly the transient and reversible nature of its modifications, introduce a new class of engineering challenges and opportunities that are distinct from those of other BNT paradigms.

\subsection{Multi-Scale and Physiologically-Aware Channel Modeling}

A defining feature of using NG-CSE to create BNTs is that the resulting entities are not passive relays or simple abiotic machines but living cells. Because the engineering is confined to the surface, the cell's internal machinery, metabolic processes, and genetic blueprint remain intact. This means each BNT is fundamentally an \emph{autonomous agent} with its own biological agenda: survival, growth, migration, proliferation, and response to its environment. This inherent agency is a critical factor that distinguishes these BNTs from all other paradigms and poses a profound modeling challenge.

Standard MC models developed for IoBNT frequently represent the transmitter as a point source releasing a pulse of molecules into a passive, unbounded medium, with the receiver being a simple absorbing or transparent sphere. This abstraction, while mathematically convenient, fails to capture the essential characteristics of surface-engineered cells. These BNTs are not merely physical objects with a finite size and dynamic membrane; they are active agents. Their agential behavior, such as motility, apoptosis, or metabolic shifts in response to local cues, cannot be entirely programmed by the IoBNT designer and can introduce a significant source of unpredictability into the network. While multi-scale modeling is an established concept in computational biology for linking molecular-level details to tissue-level phenomena, these approaches have not yet been systematically developed to create a complete communication channel model that accounts for the agency of the nodes themselves.

A paradigm shift is required to develop a new class of \emph{multi-scale channel models} that explicitly embrace and incorporate the agential nature of living BNTs. Such models must integrate phenomena across different scales of organization, from the nanoscale components on the cell surface to the emergent behavior of the entire BNT population. (i) \emph{Nanoscale (Component Level)}: At the most fundamental level, the model must incorporate the specific kinetics and efficiencies of the engineered surface elements. For instance, if communication relies on aptamer-based sensors, the model must include the binding and unbinding rate constants of the aptamers. If actuation is achieved via photothermal nanoparticles, the model must capture their light absorption cross-section and heat conversion efficiency. These nanoscale properties define the raw input/output capabilities of the BNT's engineered interface. (ii) \emph{Microscale (Cellular Level)}: At the cellular level, the model must account for the living BNT's physical presence and its dynamic nature. This includes its geometry, the lateral diffusion of surface-anchored components on the membrane, and the natural turnover rate of membrane proteins and lipids. This turnover, a direct consequence of the cell's living state, effectively defines the functional lifetime and reliability of the engineered interface, a crucial parameter for any long-term deployment. (iii) \emph{Population Scale}: To capture the crucial impact of cellular agency, the multi-scale framework must incorporate \emph{Agent-Based Modeling (ABM)}. In an ABM, each BNT can be simulated as an individual, autonomous agent with a set of internal states and behavioral rules \cite{soheilypour2018agent}. This bottom-up approach is uniquely suited to studying how complex, emergent, population-level phenomena arise from simple, local interactions. Established ABM frameworks like BSim, Morpheus, or CompuCell3D provide powerful tools to study such models \cite{matyjaszkiewicz2017bsim, swat2012multi}.

\subsection{Designing for Agency}

The recognition that surface-engineered BNTs are autonomous agents is also a fundamental application and network design challenge. Treating cellular agency as a source of noise or an obstacle to be suppressed is a limited view. The true engineering opportunity lies in designing IoBNT systems that explicitly incorporate and exploit these innate biological behaviors. This requires a shift in thinking from imposing control onto cells to designing systems that intelligently exploit their pre-existing capabilities.

The potential range of IoBNT applications is determined by what BNTs can be programmed to do. However, if the complex, built-in behaviors of living cells, such as motility, chemotaxis, and collective organization, can be harnessed, more sophisticated applications can be unlocked. The engineering challenge is to develop strategies that can reliably guide and co-opt these natural functions for a desired IoBNT application. Future IoBNT applications should be designed around programming the emergent, collective behavior of BNT swarms, rather than just the actions of individual nodes. For example, instead of commanding a single BNT to release a drug, an application could be designed to create a therapeutic, self-organizing biofilm of surface engineered living BNTs at a wound site. 

Conventional network protocols (e.g., for routing, medium access) are designed for predictable, obedient nodes that follow instructions precisely. Living BNTs violate this assumption, as their decision to relay a message, move to a new location, or even remain functional is governed by their own internal state and local environment. Therefore, new agency-aware network protocols are required that treat cellular agency as a primary routing metric. For example, a routing algorithm could make decisions based not only on the physical distance between nodes but also on their predicted behavior. It might favor a path through a region with high nutrient levels to ensure BNTs acting as relays have sufficient metabolic energy. It could also incorporate the chemotactic models described above, calculating the most biologically efficient path that a motile BNT is likely to follow.

\subsection{Managing the BNT Lifecycle - The Temporality Challenge}

A defining characteristic of NG-CSE is that the engineered modifications are not permanent. Unlike genetically encoded functions, externally attached components are subject to membrane turnover, lateral diffusion, endocytic internalization, and mechanical shedding, leading to a finite functional lifetime. This transience, while beneficial for safety and reversibility, poses a fundamental engineering challenge for the reliability and predictability of an IoBNT with NG-CSE-enabled living BNTs.

Current molecular communication models largely assume that the transmitter and receiver nodes are static in their functional capabilities over the operational timescale. For surface-engineered BNTs, this assumption is invalid. The communication interface itself, i.e., the collection of surface-tethered sensors, actuators, and logic gates, degrades over time. This means the key parameters of a communication channel, such as transmitter gain, receiver sensitivity, and noise characteristics, are not fixed but are functions of time, decaying as the surface is remodeled by the living cell.

A critical theoretical direction is to develop a new class of \emph{BNT lifetime models}. These models must go beyond simple exponential decay and capture the complex stochastic processes governing the retention of surface-tethered components. For instance, what is the probability distribution for the number of functional receptors remaining on a cell surface after $t$ hours in a high-shear-stress environment like the bloodstream? Such models, informed by experimental data, would provide a quantitative foundation for predicting the functional lifespan of a BNT based on its specific engineering method (e.g., hydrophobic insertion vs. covalent linkage).

\subsection{In-Situ Reconfiguration for Adaptive IoBNT Architectures}

The transient nature of NG-CSE can be transformed from a liability into an important feature, which is the capacity for dynamic, \emph{in-situ} reconfiguration. As surface modifications naturally decay or can be shed, they can also be replaced or updated while the BNT is operating within the body. This opens the possibility of creating adaptive IoBNT systems where BNTs can be reprogrammed on-the-fly to perform new tasks, respond to changes in the environment, or have their functionality refreshed, effectively extending the network's operational lifetime.

Engineering on-demand surface modulation requires developing the core technologies to remotely and selectively erase, write, and re-write functional components on a cell's surface post-deployment. This is a formidable engineering challenge that requires new materials, new network architectures, and new control strategies that operate at the bio-cyber interface.

A key enabling technology is the use of linkers that form reversible bonds. Dynamic covalent chemistry (DCC) offers a rich toolbox of reactions (e.g., imine or boronate ester formation) where bonds can be formed and broken under specific, mild conditions like a change in local pH or redox potential. A BNT could be engineered to shed its payload in the acidic tumor microenvironment and then become receptive to a new functional module. An even more powerful approach can involve stimuli-responsive materials, where external physical signals act as the trigger. For example, functional nanoparticles could be attached via polymer linkers that cleave upon exposure to a specific wavelength of non-invasive near-infrared (NIR) light, allowing an external operator to remotely trigger the release of an old function and enable the attachment of a new one.

A visionary but compelling future direction is to design the IoBNT architecture to be self-modifying. This involves creating specialized service or depot nodes within the network whose purpose is to re-engineer other BNTs \emph{in situ}. These could be stationary nodes that, upon receiving a command from the external network, release a pulse of chemical reagents or reprogramming packages (e.g., lipid-anchored DNA logic gates) into the local environment. These packages would then diffuse to nearby BNTs and execute a surface modification protocol, such as installing a new type of sensor. This creates a new layer in the network stack of IoBNT: a \emph{reconfiguration layer} responsible for maintaining and adapting the physical hardware of the IoBNT.

\section{Conclusion}
The realization of the IoBNT vision, i.e., seamlessly interfacing the cyber domain with the biochemical intricacies of life, demands the development of BNTs that are simultaneously programmable, biocompatible, and capable of autonomous operation in complex biological environments. While existing paradigms based on nanomaterials, passive agents, and genetic engineering face significant limitations due to toxicity, lack of autonomy, or the profound risks and burdens of genomic modification, a new approach is imperative. This paper advocates for a fourth paradigm: the transient hijacking of living cells through NG-CSE. By transforming the cell membrane into a dynamic, programmable interface decorated with synthetic molecular machinery, NG-CSE uniquely harnesses the innate agency, robustness, and biocompatibility of living cells while providing the programmability of synthetic nanotechnology. This strategy bridges the synthetic and biological worlds, unlocking unprecedented opportunities for IoBNT, including the creation of dynamic network topologies, the implementation of cell-surface logic processing, and the establishment of sophisticated bio-cyber interfaces. As demonstrated through envisioned applications such as continuous liquid biopsy using circulating sentinels and rationally designed \emph{in vitro} bio-computers, NG-CSE enables IoBNT architectures that leverage cellular agency as a core design feature. Realizing this transformative potential, however, requires addressing significant interdisciplinary challenges, including the development of multi-scale communication models that account for cellular autonomy, the design of agency-aware network protocols, and strategies for managing the transient lifecycle and enabling in-situ reconfiguration of these dynamic nodes. Despite these challenges, NG-CSE presents a flexible and high-potential pathway for the future of IoBNT, promising to deliver intelligent, adaptive, and seamlessly integrated bio-cyber systems.

\section*{Acknowledgment}
The authors thank Deniz Canim for contributions during the initiation phase of this study.

\bibliographystyle{IEEEtran}
\bibliography{cellsurface}

\begin{thebibliography}{10}
\providecommand{\url}[1]{#1}
\csname url@samestyle\endcsname
\providecommand{\newblock}{\relax}
\providecommand{\bibinfo}[2]{#2}
\providecommand{\BIBentrySTDinterwordspacing}{\spaceskip=0pt\relax}
\providecommand{\BIBentryALTinterwordstretchfactor}{4}
\providecommand{\BIBentryALTinterwordspacing}{\spaceskip=\fontdimen2\font plus
\BIBentryALTinterwordstretchfactor\fontdimen3\font minus
  \fontdimen4\font\relax}
\providecommand{\BIBforeignlanguage}[2]{{%
\expandafter\ifx\csname l@#1\endcsname\relax
\typeout{** WARNING: IEEEtran.bst: No hyphenation pattern has been}%
\typeout{** loaded for the language `#1'. Using the pattern for}%
\typeout{** the default language instead.}%
\else
\language=\csname l@#1\endcsname
\fi
#2}}
\providecommand{\BIBdecl}{\relax}
\BIBdecl

\bibitem{akyildiz2015internet}
I.~F. Akyildiz, M.~Pierobon, S.~Balasubramaniam, and Y.~Koucheryavy, ``The
  internet of bio-nano things,'' \emph{IEEE Communications Magazine}, vol.~53,
  no.~3, pp. 32--40, 2015.

\bibitem{kuscu2021internet}
M.~Kuscu and B.~D. Unluturk, ``Internet of bio-nano things: A review of
  applications, enabling technologies and key challenges,'' \emph{arXiv
  preprint arXiv:2112.09249}, 2021.

\bibitem{akan2017fundamentals}
O.~B. Akan, H.~Ramezani, T.~Khan, N.~A. Abbasi, and M.~Kuscu, ``Fundamentals of
  molecular information and communication science,'' \emph{Proc. IEEE}, vol.
  105, no.~2, pp. 306--318, 2017.

\bibitem{sun2025sensing}
Y.~Sun, K.~Wu, D.~Du, and Y.~Chen, ``Sensing by seeing: In vivo detection and
  external transmission of blood viscosity through the internet of bio-nano
  things,'' \emph{IEEE Sensors Journal}, 2025.

\bibitem{civas2023graphene}
M.~Civas, M.~Kuscu, O.~Cetinkaya, B.~E. Ortlek, and O.~B. Akan, ``Graphene and
  related materials for the internet of bio-nano things,'' \emph{APL
  Materials}, vol.~11, no.~8, 2023.

\bibitem{kuscu2021fabrication}
M.~Kuscu, H.~Ramezani, E.~Dinc, S.~Akhavan, and O.~B. Akan, ``Fabrication and
  microfluidic analysis of graphene-based molecular communication receiver for
  internet of nano things (iont),'' \emph{Scientific reports}, vol.~11, no.~1,
  p. 19600, 2021.

\bibitem{kuscu2015internet}
M.~Kuscu and O.~B. Akan, ``The internet of molecular things based on fret,''
  \emph{IEEE Internet of Things Journal}, vol.~3, no.~1, pp. 4--17, 2015.

\bibitem{von2011nanoparticles}
G.~Von~Maltzahn, J.-H. Park, K.~Y. Lin, N.~Singh, C.~Schw{\"o}ppe, R.~Mesters,
  W.~E. Berdel, E.~Ruoslahti, M.~J. Sailor, and S.~N. Bhatia, ``Nanoparticles
  that communicate in vivo to amplify tumour targeting,'' \emph{Nature
  materials}, vol.~10, no.~7, pp. 545--552, 2011.

\bibitem{koca2024bacterial}
B.~Y. Koca and O.~B. Akan, ``Bacterial communications and computing in internet
  of everything (ioe),'' \emph{IEEE Communications Surveys \& Tutorials}, 2024.

\bibitem{unluturk2015genetically}
B.~D. Unluturk, A.~O. Bicen, and I.~F. Akyildiz, ``Genetically engineered
  bacteria-based biotransceivers for molecular communication,'' \emph{IEEE
  Transactions on Communications}, vol.~63, no.~4, pp. 1271--1281, 2015.

\bibitem{wu2016metabolic}
G.~Wu, Q.~Yan, J.~A. Jones, Y.~J. Tang, S.~S. Fong, and M.~A. Koffas,
  ``Metabolic burden: cornerstones in synthetic biology and metabolic
  engineering applications,'' \emph{Trends in biotechnology}, vol.~34, no.~8,
  pp. 652--664, 2016.

\bibitem{grupi2019interfacing}
A.~Grupi, I.~Ashur, N.~Degani-Katzav, S.~Yudovich, Z.~Shapira, A.~Marzouq,
  L.~Morgenstein, Y.~Mandel, and S.~Weiss, ``Interfacing the cell with
  “biomimetic membrane proteins”,'' \emph{Small}, vol.~15, no.~52, p.
  1903006, 2019.

\bibitem{abbina2017surface}
S.~Abbina, E.~M. Siren, H.~Moon, and J.~N. Kizhakkedathu, ``Surface engineering
  for cell-based therapies: techniques for manipulating mammalian cell
  surfaces,'' \emph{ACS Biomaterials Science \& Engineering}, vol.~4, no.~11,
  pp. 3658--3677, 2017.

\bibitem{babar2024sustainable}
A.~Z. Babar and O.~B. Akan, ``Sustainable and precision agriculture with the
  internet of everything (ioe),'' \emph{arXiv preprint arXiv:2404.06341}, 2024.

\bibitem{kuscu2019transmitter}
M.~Kuscu, E.~Dinc, B.~A. Bilgin, H.~Ramezani, and O.~B. Akan, ``Transmitter and
  receiver architectures for molecular communications: A survey on physical
  design with modulation, coding, and detection techniques,'' \emph{Proceedings
  of the IEEE}, vol. 107, no.~7, pp. 1302--1341, 2019.

\bibitem{jamali2019channel}
V.~Jamali, A.~Ahmadzadeh, W.~Wicke, A.~Noel, and R.~Schober, ``Channel modeling
  for diffusive molecular communication—a tutorial review,''
  \emph{Proceedings of the IEEE}, vol. 107, no.~7, pp. 1256--1301, 2019.

\bibitem{zafar2021systematic}
S.~Zafar, M.~Nazir, T.~Bakhshi, H.~A. Khattak, S.~Khan, M.~Bilal, K.-K.~R.
  Choo, K.-S. Kwak, and A.~Sabah, ``A systematic review of bio-cyber interface
  technologies and security issues for internet of bio-nano things,''
  \emph{IEEE Access}, vol.~9, pp. 93\,529--93\,566, 2021.

\bibitem{civas2021universal}
M.~Civas, O.~Cetinkaya, M.~Kuscu, and O.~B. Akan, ``Universal transceivers:
  Opportunities and future directions for the internet of everything (ioe),''
  \emph{Frontiers in Communications and Networks}, vol.~2, p. 733664, 2021.

\bibitem{schroeder2018carbon}
V.~Schroeder, S.~Savagatrup, M.~He, S.~Lin, and T.~M. Swager, ``Carbon nanotube
  chemical sensors,'' \emph{Chemical reviews}, vol. 119, no.~1, pp. 599--663,
  2018.

\bibitem{erkoc2019mobile}
P.~Erkoc, I.~C. Yasa, H.~Ceylan, O.~Yasa, Y.~Alapan, and M.~Sitti, ``Mobile
  microrobots for active therapeutic delivery,'' \emph{Advanced Therapeutics},
  vol.~2, no.~1, p. 1800064, 2019.

\bibitem{ma2017bio}
X.~Ma and S.~S{\'a}nchez, ``Bio-catalytic mesoporous janus nano-motors powered
  by catalase enzyme,'' \emph{Tetrahedron}, vol.~73, no.~33, pp. 4883--4886,
  2017.

\bibitem{ceylan20193d}
H.~Ceylan, I.~C. Yasa, O.~Yasa, A.~F. Tabak, J.~Giltinan, and M.~Sitti,
  ``3d-printed biodegradable microswimmer for theranostic cargo delivery and
  release,'' \emph{ACS nano}, vol.~13, no.~3, pp. 3353--3362, 2019.

\bibitem{llopis2017interactive}
A.~Llopis-Lorente, P.~D{\'\i}ez, A.~S{\'a}nchez, M.~D. Marcos, F.~Sancen{\'o}n,
  P.~Mart{\'\i}nez-Ruiz, R.~Villalonga, and R.~Mart{\'\i}nez-M{\'a}{\~n}ez,
  ``Interactive models of communication at the nanoscale using nanoparticles
  that talk to one another,'' \emph{Nature communications}, vol.~8, no.~1, p.
  15511, 2017.

\bibitem{kuscu2015fluorescent}
M.~Kuscu, A.~Kiraz, and O.~B. Akan, ``Fluorescent molecules as transceiver
  nanoantennas: The first practical and high-rate information transfer over a
  nanoscale communication channel based on fret,'' \emph{Scientific reports},
  vol.~5, no.~1, p. 7831, 2015.

\bibitem{zhu2024synthetic}
B.~Zhu, H.~Yin, D.~Zhang, M.~Zhang, X.~Chao, L.~Scimeca, and M.-R. Wu,
  ``Synthetic biology approaches for improving the specificity and efficacy of
  cancer immunotherapy,'' \emph{Cellular \& Molecular Immunology}, vol.~21,
  no.~5, pp. 436--447, 2024.

\bibitem{sezgen2021multiscale}
O.~F. Sezgen, O.~Altan, A.~Bilir, M.~G. Durmaz, N.~Haciosmanoglu, B.~Camli,
  Z.~C.~C. Ozdil, A.~E. Pusane, A.~D. Yalcinkaya, U.~O.~S. Seker \emph{et~al.},
  ``A multiscale communications system based on engineered bacteria.''
  \emph{IEEE Commun. Mag.}, vol.~59, no.~5, pp. 62--67, 2021.

\bibitem{zalatan2024engineering}
J.~G. Zalatan, L.~Petrini, and R.~Geiger, ``Engineering bacteria for cancer
  immunotherapy,'' \emph{Current opinion in biotechnology}, vol.~85, p. 103061,
  2024.

\bibitem{wu2024cell}
X.~Wu, J.-J. Hu, and J.~Yoon, ``Cell membrane as a promising therapeutic
  target: from materials design to biomedical applications,'' \emph{Angewandte
  Chemie International Edition}, vol.~63, no.~18, p. e202400249, 2024.

\bibitem{ma2021functional}
S.~Ma, Y.~Xu, and W.~Song, ``Functional bionanomaterials for cell surface
  engineering in cancer immunotherapy,'' \emph{APL bioengineering}, vol.~5,
  no.~2, 2021.

\bibitem{park2018engineering}
J.~Park, B.~Andrade, Y.~Seo, M.-J. Kim, S.~C. Zimmerman, and H.~Kong,
  ``Engineering the surface of therapeutic “living” cells,'' \emph{Chemical
  reviews}, vol. 118, no.~4, pp. 1664--1690, 2018.

\bibitem{sun2017surface}
X.~Sun, X.~Han, L.~Xu, M.~Gao, J.~Xu, R.~Yang, and Z.~Liu,
  ``Surface-engineering of red blood cells as artificial antigen presenting
  cells promising for cancer immunotherapy,'' \emph{Small}, vol.~13, no.~40, p.
  1701864, 2017.

\bibitem{suh2019nanoscale}
S.~Suh, A.~Jo, M.~A. Traore, Y.~Zhan, S.~L. Coutermarsh-Ott, V.~M.
  Ringel-Scaia, I.~C. Allen, R.~M. Davis, and B.~Behkam, ``Nanoscale
  bacteria-enabled autonomous drug delivery system (nanobeads) enhances
  intratumoral transport of nanomedicine,'' \emph{Advanced Science}, vol.~6,
  no.~3, p. 1801309, 2019.

\bibitem{altman2013modifying}
M.~O. Altman, Y.~M. Chang, X.~Xiong, and W.~Tan, ``Modifying cellular
  properties using artificial aptamer-lipid receptors,'' \emph{Scientific
  reports}, vol.~3, no.~1, pp. 1--6, 2013.

\bibitem{ueki2017nongenetic}
R.~Ueki, S.~Atsuta, A.~Ueki, and S.~Sando, ``Nongenetic reprogramming of the
  ligand specificity of growth factor receptors by bispecific dna aptamers,''
  \emph{Journal of the American Chemical Society}, vol. 139, no.~19, pp.
  6554--6557, 2017.

\bibitem{nan2025advancements}
H.~Nan, M.~Cai, Y.~Wang, H.-H. Wang, and Z.~Nie, ``Advancements in dna-driven
  precision modulation of cell surface receptor for programmable cellular
  functions,'' \emph{Advanced Science}, p. e05073, 2025.

\bibitem{feng2021recent}
L.~Feng, J.~Li, J.~Sun, L.~Wang, C.~Fan, and J.~Shen, ``Recent advances of dna
  nanostructure-based cell membrane engineering,'' \emph{Advanced Healthcare
  Materials}, vol.~10, no.~6, p. 2001718, 2021.

\bibitem{guo2022manipulation}
Z.~Guo, L.~Zhang, Q.~Yang, R.~Peng, X.~Yuan, L.~Xu, Z.~Wang, F.~Chen, H.~Huang,
  Q.~Liu \emph{et~al.}, ``Manipulation of multiple cell--cell interactions by
  tunable dna scaffold networks,'' \emph{Angewandte Chemie International
  Edition}, vol.~61, no.~7, p. e202111151, 2022.

\bibitem{dobson2008remote}
J.~Dobson, ``Remote control of cellular behaviour with magnetic
  nanoparticles,'' \emph{Nature nanotechnology}, vol.~3, no.~3, pp. 139--143,
  2008.

\bibitem{zhao2021surface}
S.~Zhao, J.~Duan, Y.~Lou, R.~Gao, S.~Yang, P.~Wang, C.~Wang, L.~Han, M.~Li,
  C.~Ma \emph{et~al.}, ``Surface specifically modified nk-92 cells with cd56
  antibody conjugated superparamagnetic fe 3 o 4 nanoparticles for magnetic
  targeting immunotherapy of solid tumors,'' \emph{Nanoscale}, vol.~13, no.~45,
  pp. 19\,109--19\,122, 2021.

\bibitem{wang2024cell}
Y.~Wang, J.~Shi, M.~Xin, A.~R. Kahkoska, J.~Wang, and Z.~Gu, ``Cell--drug
  conjugates,'' \emph{Nature biomedical engineering}, vol.~8, no.~11, pp.
  1347--1365, 2024.

\bibitem{stephan2011enhancing}
M.~T. Stephan and D.~J. Irvine, ``Enhancing cell therapies from the outside in:
  cell surface engineering using synthetic nanomaterials,'' \emph{Nano today},
  vol.~6, no.~3, pp. 309--325, 2011.

\bibitem{lee2023nanoparticles}
S.~Lee, M.~Jiao, Z.~Zhang, and Y.~Yu, ``Nanoparticles for interrogation of cell
  signaling,'' \emph{Annual Review of Analytical Chemistry}, vol.~16, no.~1,
  pp. 333--351, 2023.

\bibitem{ali2012cell}
M.~M. Ali, D.-K. Kang, K.~Tsang, M.~Fu, J.~M. Karp, and W.~Zhao, ``Cell-surface
  sensors: lighting the cellular environment,'' \emph{Wiley Interdisciplinary
  Reviews: Nanomedicine and Nanobiotechnology}, vol.~4, no.~5, pp. 547--561,
  2012.

\bibitem{Ghezzi2021-hd}
M.~Ghezzi, S.~Pescina, C.~Padula, P.~Santi, E.~Del~Favero, L.~Cant{\`u}, and
  S.~Nicoli, ``\BIBforeignlanguage{en}{Polymeric micelles in drug delivery: An
  insight of the techniques for their characterization and assessment in
  biorelevant conditions},'' \emph{\BIBforeignlanguage{en}{J. Control.
  Release}}, vol. 332, pp. 312--336, Apr. 2021.

\bibitem{wang2015non}
Q.~Wang, H.~Cheng, H.~Peng, H.~Zhou, P.~Y. Li, and R.~Langer, ``Non-genetic
  engineering of cells for drug delivery and cell-based therapy,''
  \emph{Advanced drug delivery reviews}, vol.~91, pp. 125--140, 2015.

\bibitem{lee2018cell}
D.~Y. Lee, B.-H. Cha, M.~Jung, A.~S. Kim, D.~A. Bull, and Y.-W. Won, ``Cell
  surface engineering and application in cell delivery to heart diseases,''
  \emph{Journal of Biological Engineering}, vol.~12, no.~1, pp. 1--11, 2018.

\bibitem{niu2017engineering}
J.~Niu, D.~J. Lunn, A.~Pusuluri, J.~I. Yoo, M.~A. O'Malley, S.~Mitragotri,
  H.~T. Soh, and C.~J. Hawker, ``Engineering live cell surfaces with functional
  polymers via cytocompatible controlled radical polymerization,'' \emph{Nature
  Chemistry}, vol.~9, no.~6, pp. 537--545, 2017.

\bibitem{zhong2023site}
Y.~Zhong, L.~Xu, C.~Yang, L.~Xu, G.~Wang, Y.~Guo, S.~Cheng, X.~Tian, C.~Wang,
  R.~Xie \emph{et~al.}, ``Site-selected in situ polymerization for living cell
  surface engineering,'' \emph{Nature Communications}, vol.~14, no.~1, p. 7285,
  2023.

\bibitem{henry2020kodecytes}
S.~Henry, ``Kodecytes: modifying the surface of red blood cells,'' \emph{ISBT
  Science Series}, vol.~15, no.~3, pp. 303--309, 2020.

\bibitem{ma2023advances}
X.~Ma, J.~Jiang, X.~An, W.~Zu, C.~Ma, Z.~Zhang, Y.~Lu, L.~Zhao, and L.~Wang,
  ``Advances in research based on antibody-cell conjugation,'' \emph{Frontiers
  in Immunology}, vol.~14, p. 1310130, 2023.

\bibitem{ye2023nongenetic}
T.~Ye, X.~Liu, X.~Zhong, R.~Yan, and P.~Shi, ``Nongenetic surface engineering
  of mesenchymal stromal cells with polyvalent antibodies to enhance targeting
  efficiency,'' \emph{Nature Communications}, vol.~14, no.~1, p. 5806, 2023.

\bibitem{park2024ex}
H.~W. Park, C.~E. Lee, S.~Kim, W.-J. Jeong, and K.~Kim, ``Ex vivo peptide
  decoration strategies on stem cell surfaces for augmenting endothelium
  interaction,'' \emph{Tissue Engineering Part B: Reviews}, vol.~30, no.~3, pp.
  327--339, 2024.

\bibitem{wu2023chemically}
H.~Wu, L.~Zheng, N.~Ling, L.~Zheng, Y.~Du, Q.~Zhang, Y.~Liu, W.~Tan, and
  L.~Qiu, ``Chemically synthetic membrane receptors establish cells with
  artificial sense-and-respond signaling pathways,'' \emph{Journal of the
  American Chemical Society}, vol. 145, no.~4, pp. 2315--2321, 2023.

\bibitem{almeida2023cell}
J.~Almeida-Pinto, M.~R. Lagarto, P.~Lavrador, J.~F. Mano, and V.~M. Gaspar,
  ``Cell surface engineering tools for programming living assemblies,''
  \emph{Advanced Science}, vol.~10, no.~34, p. 2304040, 2023.

\bibitem{roy2020nongenetic}
S.~Roy, J.~N. Cha, and A.~P. Goodwin, ``Nongenetic bioconjugation strategies
  for modifying cell membranes and membrane proteins: a review,''
  \emph{Bioconjugate chemistry}, vol.~31, no.~11, pp. 2465--2475, 2020.

\bibitem{csizmar2018programming}
C.~M. Csizmar, J.~R. Petersburg, and C.~R. Wagner, ``Programming cell-cell
  interactions through non-genetic membrane engineering,'' \emph{Cell chemical
  biology}, vol.~25, no.~8, pp. 931--940, 2018.

\bibitem{liu2019advances}
L.~Liu, H.~He, and J.~Liu, ``Advances on non-genetic cell membrane engineering
  for biomedical applications,'' \emph{Polymers}, vol.~11, no.~12, p. 2017,
  2019.

\bibitem{schoenit2021functionalization}
A.~Schoenit, E.~A. Cavalcanti-Adam, and K.~G{\"o}pfrich, ``Functionalization of
  cellular membranes with dna nanotechnology,'' \emph{Trends in Biotechnology},
  vol.~39, no.~11, pp. 1208--1220, 2021.

\bibitem{jia2022recent}
H.-R. Jia, Z.~Zhang, X.~Fang, M.~Jiang, M.~Chen, S.~Chen, K.~Gu, Z.~Luo, F.-G.
  Wu, and W.~Tan, ``Recent advances of cell surface modification based on
  aptamers,'' \emph{Materials Today Nano}, vol.~18, p. 100188, 2022.

\bibitem{li2018dna}
H.~Li, M.~Wang, T.~Shi, S.~Yang, J.~Zhang, H.-H. Wang, and Z.~Nie, ``A
  dna-mediated chemically induced dimerization (d-cid) nanodevice for
  nongenetic receptor engineering to control cell behavior,'' \emph{Angewandte
  Chemie International Edition}, vol.~57, no.~32, pp. 10\,226--10\,230, 2018.

\bibitem{chen2019logic}
S.~Chen, Z.~Xu, W.~Yang, X.~Lin, J.~Li, J.~Li, and H.~Yang,
  ``Logic-gate-actuated dna-controlled receptor assembly for the programmable
  modulation of cellular signal transduction,'' \emph{Angewandte Chemie
  International Edition}, vol.~58, no.~50, pp. 18\,186--18\,190, 2019.

\bibitem{wu2023logic}
Y.~Wu, J.~Huang, H.~He, M.~Wang, G.~Yin, L.~Qi, X.~He, H.-H. Wang, and K.~Wang,
  ``Logic nanodevice-mediated receptor assembly for nongenetic regulation of
  cell behavior in tumor-like microenvironment,'' \emph{Nano Letters}, vol.~23,
  no.~5, pp. 1801--1809, 2023.

\bibitem{li2019cell}
J.~Li, K.~Xun, K.~Pei, X.~Liu, X.~Peng, Y.~Du, L.~Qiu, and W.~Tan,
  ``Cell-membrane-anchored dna nanoplatform for programming cellular
  interactions,'' \emph{Journal of the American Chemical Society}, vol. 141,
  no.~45, pp. 18\,013--18\,020, 2019.

\bibitem{ge2020programming}
Z.~Ge, J.~Liu, L.~Guo, G.~Yao, Q.~Li, L.~Wang, J.~Li, and C.~Fan, ``Programming
  cell--cell communications with engineered cell origami clusters,''
  \emph{Journal of the American Chemical Society}, vol. 142, no.~19, pp.
  8800--8808, 2020.

\bibitem{wang2023dna}
K.~Wang, Y.~Wei, X.~Xie, Q.~Li, X.~Liu, L.~Wang, J.~Li, J.~Wu, and C.~Fan,
  ``Dna-programmed stem cell niches via orthogonal extracellular vesicle--cell
  communications,'' \emph{Advanced Materials}, vol.~35, no.~45, p. 2302323,
  2023.

\bibitem{kim2018general}
H.~Kim, K.~Shin, O.~K. Park, D.~Choi, H.~D. Kim, S.~Baik, S.~H. Lee, S.-H.
  Kwon, K.~J. Yarema, J.~Hong \emph{et~al.}, ``General and facile coating of
  single cells via mild reduction,'' \emph{Journal of the American Chemical
  Society}, vol. 140, no.~4, pp. 1199--1202, 2018.

\bibitem{kim2015recent}
J.~C. Kim and G.~Tae, ``Recent advances in cell surface engineering focused on
  cell therapy,'' \emph{Bulletin of the Korean Chemical Society}, vol.~36,
  no.~1, pp. 59--65, 2015.

\bibitem{fu2024surface}
L.~Fu, Q.~He, X.~Lu, L.~Hu, H.~Qian, and P.~Pei, ``Surface engineering on
  bacteria for tumor immunotherapy: strategies and perspectives,''
  \emph{Advanced Functional Materials}, vol.~34, no.~42, p. 2405304, 2024.

\bibitem{song2024engineering}
L.~Song, Y.~Wang, Y.~Guo, S.~Bulale, M.~Zhou, F.~Yu, and L.~He, ``Engineering
  aptamers to enhance their interaction with protein target for selective
  inhibition of cell surface receptors,'' \emph{International Journal of
  Biological Macromolecules}, vol. 278, p. 134989, 2024.

\bibitem{li2023transformable}
B.~Li, Y.~Wang, and B.~Liu, ``Transformable dna nanorobots reversibly
  regulating cell membrane receptors for modulation of cellular migrations,''
  \emph{ACS nano}, vol.~17, no.~22, pp. 22\,571--22\,579, 2023.

\bibitem{su2022enhancing}
Y.~Su, X.~Chen, H.~Wang, L.~Sun, Y.~Xu, and D.~Li, ``Enhancing cell membrane
  phase separation for inhibiting cancer metastasis with a stimuli-responsive
  dna nanodevice,'' \emph{Chemical Science}, vol.~13, no.~21, pp. 6303--6308,
  2022.

\bibitem{wang2023collective}
X.~Wang, X.~Xing, S.~Lu, G.~Du, Y.~Zhang, Y.~Ren, Y.~Sun, J.~Sun, Q.~Fan,
  K.~Liu \emph{et~al.}, ``Collective cell behaviors manipulated by synthetic
  dna nanostructures,'' \emph{Fundamental Research}, vol.~3, no.~5, pp.
  809--812, 2023.

\bibitem{emiliani2022optogenetics}
V.~Emiliani, E.~Entcheva, R.~Hedrich, P.~Hegemann, K.~R. Konrad,
  C.~L{\"u}scher, M.~Mahn, Z.-H. Pan, R.~R. Sims, J.~Vierock \emph{et~al.},
  ``Optogenetics for light control of biological systems,'' \emph{Nature
  Reviews Methods Primers}, vol.~2, no.~1, p.~55, 2022.

\bibitem{collot2022advanced}
M.~Collot, S.~Pfister, and A.~S. Klymchenko, ``Advanced functional fluorescent
  probes for cell plasma membranes,'' \emph{Current Opinion in Chemical
  Biology}, vol.~69, p. 102161, 2022.

\bibitem{you2017dna}
M.~You, Y.~Lyu, D.~Han, L.~Qiu, Q.~Liu, T.~Chen, C.~Sam~Wu, L.~Peng, L.~Zhang,
  G.~Bao \emph{et~al.}, ``Dna probes for monitoring dynamic and transient
  molecular encounters on live cell membranes,'' \emph{Nature nanotechnology},
  vol.~12, no.~5, pp. 453--459, 2017.

\bibitem{zhang2023functional}
Z.~Zhang, Y.~You, M.~Ge, H.~Lin, and J.~Shi, ``Functional nanoparticle-enabled
  non-genetic neuromodulation,'' \emph{Journal of Nanobiotechnology}, vol.~21,
  no.~1, p. 319, 2023.

\bibitem{nag2020nanoparticle}
O.~K. Nag, M.~E. Muroski, D.~A. Hastman, B.~Almeida, I.~L. Medintz, A.~L.
  Huston, and J.~B. Delehanty, ``Nanoparticle-mediated visualization and
  control of cellular membrane potential: Strategies, progress, and remaining
  issues,'' \emph{Acs Nano}, vol.~14, no.~3, pp. 2659--2677, 2020.

\bibitem{huang2023nanoparticles}
Q.~Huang, W.~Zhu, X.~Gao, X.~Liu, Z.~Zhang, and B.~Xing,
  ``Nanoparticles-mediated ion channels manipulation: From their membrane
  interactions to bioapplications,'' \emph{Advanced Drug Delivery Reviews},
  vol. 195, p. 114763, 2023.

\bibitem{wang2019near}
M.~Wang, F.~He, H.~Li, S.~Yang, J.~Zhang, P.~Ghosh, H.-H. Wang, and Z.~Nie,
  ``Near-infrared light-activated dna-agonist nanodevice for nongenetically and
  remotely controlled cellular signaling and behaviors in live animals,''
  \emph{Nano letters}, vol.~19, no.~4, pp. 2603--2613, 2019.

\bibitem{liu2016nanoscale}
Z.~Liu, Y.~Liu, Y.~Chang, H.~R. Seyf, A.~Henry, A.~L. Mattheyses, K.~Yehl,
  Y.~Zhang, Z.~Huang, and K.~Salaita, ``Nanoscale optomechanical actuators for
  controlling mechanotransduction in living cells,'' \emph{Nature methods},
  vol.~13, no.~2, pp. 143--146, 2016.

\bibitem{hicks2021electric}
J.~M. Hicks, Y.-C. Yao, S.~Barber, N.~Neate, J.~A. Watts, A.~Noy, and F.~J.
  Rawson, ``Electric field induced biomimetic transmembrane electron transport
  using carbon nanotube porins,'' \emph{Small}, vol.~17, no.~32, p. 2102517,
  2021.

\bibitem{padma2023artificial}
K.~Padma and K.~Don, ``Artificial photosynthesis with gold nanostructures
  incorporation in non-photosynthetic bacteria,'' in \emph{Modern
  Nanotechnology: Volume 1: Environmental Sustainability and
  Remediation}.\hskip 1em plus 0.5em minus 0.4em\relax Springer, 2023, pp.
  669--682.

\bibitem{rakovich2010resonance}
A.~Rakovich, A.~Sukhanova, N.~Bouchonville, E.~Lukashev, V.~Oleinikov,
  M.~Artemyev, V.~Lesnyak, N.~Gaponik, M.~Molinari, M.~Troyon \emph{et~al.},
  ``Resonance energy transfer improves the biological function of
  bacteriorhodopsin within a hybrid material built from purple membranes and
  semiconductor quantum dots,'' \emph{Nano letters}, vol.~10, no.~7, pp.
  2640--2648, 2010.

\bibitem{li2007stationary}
R.~Li, C.~M. Li, H.~Bao, Q.~Bao, and V.~S. Lee, ``Stationary current generated
  from photocycle of a hybrid bacteriorhodopsin/quantum dot bionanosystem,''
  \emph{Applied Physics Letters}, vol.~91, no.~22, 2007.

\bibitem{gao2022circulating}
Q.~Gao, Q.~Zeng, Z.~Wang, C.~Li, Y.~Xu, P.~Cui, X.~Zhu, H.~Lu, G.~Wang, S.~Cai
  \emph{et~al.}, ``Circulating cell-free dna for cancer early detection,''
  \emph{The Innovation}, vol.~3, no.~4, 2022.

\bibitem{brenner2021red}
J.~S. Brenner, S.~Mitragotri, and V.~R. Muzykantov, ``Red blood cell
  hitchhiking: a novel approach for vascular delivery of nanocarriers,''
  \emph{Annual review of biomedical engineering}, vol.~23, no.~1, pp. 225--248,
  2021.

\bibitem{cai2023brain}
H.~Cai, Z.~Ao, C.~Tian, Z.~Wu, H.~Liu, J.~Tchieu, M.~Gu, K.~Mackie, and F.~Guo,
  ``Brain organoid reservoir computing for artificial intelligence,''
  \emph{Nature Electronics}, vol.~6, no.~12, pp. 1032--1039, 2023.

\bibitem{soheilypour2018agent}
M.~Soheilypour and M.~R. Mofrad, ``Agent-based modeling in molecular systems
  biology,'' \emph{BioEssays}, vol.~40, no.~7, p. 1800020, 2018.

\bibitem{matyjaszkiewicz2017bsim}
A.~Matyjaszkiewicz, G.~Fiore, F.~Annunziata, C.~S. Grierson, N.~J. Savery,
  L.~Marucci, and M.~Di~Bernardo, ``Bsim 2.0: an advanced agent-based cell
  simulator,'' \emph{ACS synthetic biology}, vol.~6, no.~10, pp. 1969--1972,
  2017.

\bibitem{swat2012multi}
M.~H. Swat, G.~L. Thomas, J.~M. Belmonte, A.~Shirinifard, D.~Hmeljak, and J.~A.
  Glazier, ``Multi-scale modeling of tissues using compucell3d,'' in
  \emph{Methods in cell biology}.\hskip 1em plus 0.5em minus 0.4em\relax
  Elsevier, 2012, vol. 110, pp. 325--366.

\end{thebibliography}
\end{document}